%% file: LFVHD_mainfile.tex
\documentclass[12pt]{article}
\usepackage{amsmath,amssymb,amsfonts,color,graphicx,cite,color}
\usepackage{latexsym}
\usepackage{epsfig,psfrag,rotating,soul}
\input paperdef

\evensidemargin \oddsidemargin
\allowdisplaybreaks

\begin{document}
\thispagestyle{empty}

\def\thefootnote{\fnsymbol{footnote}}

\begin{flushright}
\mbox{}
\end{flushright}

\vspace{0.5cm}

\begin{center}

\begin{large}
\textbf{Lepton flavor violating Higgs Boson Decays in }
\\[2ex]
\textbf{Supersymmetric High Scale Seesaw Models}
\end{large}

\vspace{1cm}

{\sc
M.E.~G{\'o}mez$^{1}$%
\footnote{email: mario.gomez@dfa.uhu.es}%
, S.~Heinemeyer$^{2,3,4}$%
\footnote{email: Sven.Heinemeyer@cern.ch}%
~and M.~Rehman$^{5}$%
\footnote{email: muhammad.rehman@riphah.edu.pk}%
 
}

\vspace*{.7cm}

{\sl
$^1$ Departamento de Ciencias Integradas, Universidad de Huelva, 21071 Huelva, Spain
\vspace*{0.1cm}

$^2$Instituto de F\'isica de Cantabria (CSIC-UC),  39005 Santander, Spain
\vspace*{0.1cm}

$^3$Instituto de F\'{\i}sica Te\'orica, (UAM/CSIC), Universidad Aut\'onoma de Madrid
Cantoblanco, E-28049 Madrid, Spain
\vspace*{0.1cm}

$^4$Campus of International Excellence UAM+CSIC, 
Cantoblanco, 28049, Madrid, Spain
\vspace*{0.1cm}

$^5$Riphah Institute of Computing and Applied Sciences, Riphah International University, 54770 Lahore, Pakistan
}

\end{center}

\vspace*{0.1cm}

\begin{abstract}
\noindent

Within the MSSM, we have evaluated the decay rates for the lepton flavour violating Higgs boson decays (LFVHD) $h \rightarrow l_i l_j$ where $l_{i,j}$ are charged leptons and $i\neq j$. This has been done  in a model independent (MI) way as well as in supersymmetric high scale seesaw models, in particular Type I see-saw model. Lepton flavour violation (LFV) is generated by 
non-diagonal entries in the mass matrix of the sleptons. In a first step we use the model independent approach where LFV (off-diagonal entries in the mass matrix) is introduced by hand while respecting the direct search constraints from the charged lepton flavor violating (cLFV) processes. In the second step we use high scale see-saw models where LFV is generated via renormalization group equations (RGE) from the grand unification scale (GUT)  down to electroweak scale. cLFV decays are the most restrictive ones and exclude a large part of the parameter space for the MI as well as the high scale see-saw scenarios. Due to very strict constraints from cLFV, it is difficult to find large corrections to LFVHD. This applies in particular to $h \rightarrow \tau \mu$ where hints of an excess have been observed. If this signal is confirmed, it could  not be explained with the models under investigation.

\end{abstract}

\def\thefootnote{\arabic{footnote}}
\setcounter{page}{0}
\setcounter{footnote}{0}

\newpage


\input{sec1_intro}

\input{sec2_LFVSUSY}
\input{sec3_observables}

\input{sec4_MI_analysis}

\input{sec5_CMSSM-SeesawI}
\input{sec6_conclu}
\vspace{-0.5em}
\subsection*{Acknowledgments}

The work of S.H.\ and M.R.\ was partially supported by CICYT (grant FPA
2013-40715-P). 
M.G., S.H.\ and M.R.\ were supported by 
the Spanish MICINN's Consolider-Ingenio 2010 Programme under grant
MultiDark CSD2009-00064. 
M.E.G.\  and M.R. acknowledges further support from the
MICINN project FPA2014-53631-C2-2-P


\newpage

\end{document}

%% file: sec1_intro.tex

\section{Introduction}
The Standard Model (SM) predicts flavor mixing in the quark sector. However, lepton flavor violation (LFV) is exactly zero due to the assumption of vanishing neutrino masses. The observation of neutrino oscillations\cite{Neutrino-Osc} certainly contradicts the SM, and also suggest the possibilty of the observation of flavour violation on the chaged sector (cLFV). However, processes such as $l_i \rightarrow l_j \gamma$, with $i\neq j$ and $l_{i, j}= e, \mu, \tau$ have not been observed yet. Even if the SM is complemented with massive non-degenerate light neutrinos, the rates for these processes are supressed by a 
factor $\Delta m_\nu^4/M_W^4$ where $\Delta m_\nu$ denotes the neutrino mass splitting and $M_W$ the $W$ boson mass. Data from neutrino oscillations implies values for  $\Delta m_\nu$ so small that the processes like BR($l_i\rightarrow l_j  \gamma$) would be out of the experimental scope.  Independently of the neutrino problem, even  the Minimal Supersymetric Standard Model (MSSM)~\cite{mssm} can predict charged LFV due to flavor mixing in the sleptons (scalar partners of the leptons) allowing prediction for these process in the experimental reach~\cite{Hall:1985dx,Borzumati:1986qx}. The same mechanism can enable LFV Higgs decays, such decays have gathered a lot of attention after CMS reported excess for the channel 
$h\rightarrow \mu \tau$ \cite{cms}.  This seems to be consistent with  the latest analysis of the  ATLAS results \cite{atlas}.  However, their significance is not large enough  and further data is needed to confirm or exclude this excess. 

The complementation of the MSSM with a mechanism to explain neutrino oscillations can relate those to corresponding cLFV effects. A  first guess would be to write down the neutrino yukawa couplings which generates neutrino masses via electroweak symmetry breaking (EWSB). However those couplings will be so small that it will be very difficult to link them to the observation of cLFV. This picture changes when these masses are explained with  a ``see-saw'' mechanism \cite{seesaw:I}, that can be implemented in different ways \cite{King:2003jb,Senjanovic:2011zz}. The most popular of these mechanisms is Type-I see-saw\cite{seesaw:I}, the small neutrino mass $m_{\nu} \approx Y_{\nu}^{2} v^{2}/M_{R}$ with $Y_{\nu}$ the neutrino yukawa coupling, $M_R$ the seesaw scale and $v$ the vacuum expectation value, is achieved with a high scale $M_R$ which can allow large values for $Y_{\nu}$.  Even with the assumption of universal soft masses at the GUT scale, the presence of  $Y_{\nu}$ in the RGE above $M_R$ can generate non trivial slepton mixings, hence relating cLFV to the neutrino problem \cite{LFVhisano,gllv,casas-ibarra,Mismatch,Antusch,Antusch2, Schwieger:1998dd,Divari:2002sq, Biggio:2010me,Figueiredo:2013tea,Chowdhury:2013jta,Krauss:2013gya,Goto:2014vga,Kersten:2014xaa,Vicente:2015cka} and GUT scenarios 
\cite{Barbieri:1995tw,Gomez:1995cv,pedro,Ellis:2016qra}. Other popular high scale seesaw mechanisms are Type II \cite{Magg:1980ut,Lazarides:1980nt} and Type III \cite{Foot-Type-III,Ma:2002pf} seesaw models. In Type II seesaw,  the heavy particle is a Higgs triplet,  whereas in Type III see-saw model, the exchanged particle 
should be a right-handed fermion triplet. At low energy, the neutrino masses are generated by a dimension 5 operator and one can not distinguish between different see-saw realizations. One common feature among these models is that the LFV effects in these models are generated by non diagonal entries (as explained above for the Type I see-saw mechanism) in the slepton mass matrix. These off-diagonal entries in the slepton mass matrix not only predict sizeable rates for the cLFV processes but can also results in the LFV decays of the Higgs boson \cite{Brignole:2003iv, Brignole:2004ah, Kanemura:2004cn, Paradisi:2005tk, Arana-Catania:2013xma, Arganda:2015uca, Aloni:2015wvn}.  While supersymmetric high scale see-saw models successfully describe the neutrino masses and mixing and predict sizeable rates for the cLFV processes, it is yet to be seen if they can also explain the CMS reported excess, which precisely is the aim of this work.

In this article we evaluate LFV Higgs decays like $h\rightarrow l_i l_j$ where $l_{i,j=e,\mu,\tau}$ are the charged leptons with $i\neq j$. For our calculations we prepared an add-on model file for \fa~\cite{feynarts,famssm} which adds LFV effects to the 
existing MSSM model file, as described in \cite{drhoLFV, Gomez:2015ila}. We carry out our numerical analysis in two frameworks. In the
first framework we study several expamples of mass spectra for the MSSM consistent with all the phenomenological constraints. Flavor mixing is generated by putting off-diagonal enteries in the slepton mass matrices by hand such that cLFV is consistent with direct experimental searches. In the second framework, we study MSSM augmented by the high scale seesaw models in particular Type~I seesaw mechanism\cite{seesaw:I} and flavor mixing is generated through RGEs as explained above.

This paper is organised in the following way: In section 2 the MSSM is presented and we introduce our definitions of the slepton basis and mass matrices. The third section is dedicated to briefly review the observables that will be studied in this paper. In the fourth section we present our numerical analysis in the MI approach for the observables of section 3. In section 5 we present our numerical analysis for the MSSM augmented by seesaw Type~I mechanism. Finally,  our conclusions can be found in section 6.

%% file: sec2_LFVSUSY.tex
\section{LFV in the MSSM}
\label{sec:model_setup}
The MSSM is the most popular SUSY extension of the SM. With the assumption of soft SUSY breaking terms we introduce a flavor mismatch for the scalar partners with respect to their corresponding leptons. Therefore, flavor violation is introduced through loops containing SUSY particles. In this section, along with the MSSM  we introduce the definitions and operational basis that will be used in the rest of the work. We use the same notation as in Refs. \cite{Arana-Catania:2013xma, Arganda:2004bz, Arganda:2015uca, Gomez:2015ila, Arana-Catania:2013nha}.
      
One can write the most general $SU(3)_{C}\times SU(2)_{L}\times
U(1)_{Y}$ gauge invariant and renormalizable R-parity conserving superpotential for the MSSM as
\begin{eqnarray}
\label{superpotential}
W_{\rm MSSM}&=&Y_e^{ij}\epsilon_{\alpha \beta} H_1^{\alpha} E_i^c  L_j^{\beta}
+ Y_{d}^{ij} \epsilon_{\alpha \beta} H_1^{\alpha} D_i^c  Q_j^{\beta}
+ Y_{u}^{ij} \epsilon_{\alpha \beta} H_2^{\alpha} U_i^c Q_j^{\beta}
\nonumber \\
&&+ \mu \epsilon_{\alpha \beta} H_1^{\alpha} H_2^{\beta}
\end{eqnarray}
where $L_i$ represents the chiral multiplet of a $SU(2)_L$ doublet
lepton, $E_i^c$ a $SU(2)_L$ singlet charged lepton, $H_1$ and $H_2$ two Higgs doublets with opposite hypercharge.
Similarly $Q$, $U$ and $D$ represent chiral multiplets of quarks of a
$SU(2)_L$ doublet and two singlets with different $U(1)_Y$ charges.
$Y_u$, $Y_d$ and $Y_e$ are the Yukawa couplings for up-type, down-type and charged leptons respectively. Three generations of leptons and quarks are assumed and thus the subscripts $i$ and $j$ run over 1 to 3. The symbol $\epsilon_{\alpha
\beta}$ is an anti-symmetric tensor with $\epsilon_{12}=1$.  

The general set-up for the soft SUSY-breaking
parameters is given by~\cite{mssm}
\begin{eqnarray}
\label{softbreaking}
-\cL_{\rm soft}&=&(m_{\tilde Q}^2)_i^j {\tilde q}_{L}^{\dagger i}
{\tilde q}_{Lj}
+(m_{\tilde u}^2)^i_j {\tilde u}_{Ri}^* {\tilde u}_{R}^j
+(m_{\tilde d}^2)^i_j {\tilde d}_{Ri}^* {\tilde d}_{R}^j
\nonumber \\
& &+(m_{\tilde L}^2)_i^j {\tilde l}_{L}^{\dagger i}{\tilde l}_{Lj}
+(m_{\tilde e}^2)^i_j {\tilde e}_{Ri}^* {\tilde e}_{R}^j
\nonumber \\
& &+{\tilde m}^2_{1}h_1^{\dagger} h_1
+{\tilde m}^2_{2}h_2^{\dagger} h_2
+(B \mu h_1 h_2
+ {\rm h.c.})
\nonumber \\
& &+ ( A_d^{ij}h_1 {\tilde d}_{Ri}^*{\tilde q}_{Lj}
+A_u^{ij}h_2 {\tilde u}_{Ri}^*{\tilde q}_{Lj}
+A_e^{ij}h_1 {\tilde e}_{Ri}^*{\tilde l}_{Lj}
\nonumber \\
& & +\frac{1}{2}M_1 {\tilde B}_L^0 {\tilde B}_L^0
+\frac{1}{2}M_2 {\tilde W}_L^a {\tilde W}_L^a
+\frac{1}{2}M_3 {\tilde G}^a {\tilde G}^a + {\rm h.c.}).
\end{eqnarray}
Here $m_{\tilde Q}^2$ and $m_{\tilde L}^2$ are $3 \times 3$
matrices in family space (with $i,j$ being the
generation indeces) for the soft masses of the
left handed squark ${\tilde q}_{L}$ and slepton ${\tilde l}_{L}$
$SU(2)$ doublets, respectively. $m_{\tilde u}^2$, $m_{\tilde d}^2$ and
$m_{\tilde e}^2$ contain the soft masses for right handed up-type squark
${\tilde u}_{R}$,  down-type squarks ${\tilde d}_{R}$ and charged
slepton ${\tilde e}_{R}$ $SU(2)$ singlets, respectively. $A_u$, $A_d$
and $A_e$ are the $3 \times 3$ matrices for the trilinear
couplings for up-type squarks, down-type 
squarks and charged slepton, respectively.   
${\tilde m}_1$ and ${\tilde m}_2$ are the soft
masses of the Higgs sector. In the last line $M_1$, $M_2$ and $M_3$
define the bino, wino  and gluino mass terms, respectively.

The most general hypothesis for flavor mixing in sleptons assumes a mass matrix that
is not diagonal in flavor space. 
In the charged slepton sector we have a $6 \times 6$
mass matrix, based on the corresponding six electroweak interaction
eigenstates,  ${\tilde L}_{L,R}$ with $L=e, \mu, \tau$ for charged sleptons.
For the sneutrinos 
we have a $3 \times 3$ mass matrix, since within the MSSM,  
we have only three electroweak interaction eigenstates, ${\tilde \nu}_{L}$
with $\nu=\nu_e, \nu_\mu, \nu_\tau$. 

The non-diagonal entries in this $6 \times 6$ general matrix for sleptons
can be described in terms of a set of 
dimensionless parameters $\deFABij$ ($F=L,E; A,B=L,R$; $i,j=1,2,3$, 
$i \neq j$) where  $F$ identifies the slepton type, $L,R$ refer to the 
``left-'' and ``right-handed'' SUSY partners of the corresponding
fermionic degrees of freedom, and $i,j$
indexes run over the three generations. 

One usually writes the $6\times 6$ non-diagonal mass matrices,  
${\mathcal M}_{\tilde l}^2$ referred to the
Super-PMNS basis, being ordered as $(\SelL, \SmuL, \StauL, \SelR, \SmuR,
\StauR)$, and write them in terms of left- and right-handed blocks
 $M^2_{\tilde l \, AB}$ ($A,B=L,R$),
which are non-diagonal $3\times 3$ matrices, 
\begin{equation}
{\mathcal M}_{\tilde l}^2 =\left( \begin{array}{cc}
M^2_{\tilde l \, LL} & M^2_{\tilde l \, LR} \\[.3em]
M_{\tilde l \, LR}^{2 \, \dagger} & M^2_{\tilde l \,RR}
\end{array} \right),
\label{eq:slep-6x6}
\end{equation} 
 where:
 \begin{alignat}{5}
M_{\tilde l \, LL \, ij}^2 
  = &  m_{\tilde L \, ij}^2 + \left( m_{l_i}^2
     + (-\edz + \sw^2 ) \MZ^2 \cos 2\beta \right) \delta_{ij},  \notag\\
 M^2_{\tilde l \, RR \, ij}
  = &  m_{\tilde E \, ij}^2 + \left( m_{l_i}^2
     -\sw^2 \MZ^2 \cos 2\beta \right) \delta_{ij} \notag, \\
  M^2_{\tilde l \, LR \, ij}
  = &  \left< h_1^0 \right> {\cal A}_{ij}^l- m_{l_{i}} \mu \tb \, \delta_{ij},
\label{eq:slep-matrix}
\end{alignat}
with, $i,j=1,2,3$, $\sw^2 = 1 - \MW^2/\MZ^2$ with $M_{Z,W}$ denote the $Z$~and $W$~boson masses and $(m_{l_1},m_{l_2},
m_{l_3})=(m_e,m_\mu,m_\tau)$ are the lepton masses. $\mu$ is the
Higgsino mass term and $\tb = v_2/v_1$
with  $v_1=\left< h_1^0 \right>$ and $v_2=\left< h_2^0
\right>$ being the two vacuum expectation values of the corresponding
neutral Higgs boson in the Higgs $SU(2)_L$ doublets, 
$h_1= (h^0_1\,\,\, h^-_1)$ and $h_2= (h^+_2 \,\,\,h^0_2)$.

It should be noted that the non-diagonality in flavor in the MSSM comes
exclusively from the soft SUSY-breaking parameters, that could be
non-vanishing for $i \neq j$, namely: the masses $m_{\tilde L \, ij}$ 
for the sfermion $SU(2)$ doublets, the masses $m_{\tilde E \, ij}$ for the sfermion $SU(2)$ 
singlets and the trilinear couplings, ${\cal A}_{ij}^f$.   

In the sneutrino sector there is, correspondingly, a one-block $3\times
3$ mass matrix, that is referred to the $(\tinu_{eL}, \tinu_{\mu L},
\tinu_{\tau L})$ electroweak interaction basis: 
\begin{equation}
{\mathcal M}_{\tilde \nu}^2 =\left( \begin{array}{c}
M^2_{\tilde \nu \, LL}   
\end{array} \right),
\label{eq:sneu-3x3}
\end{equation} 
 where:
\begin{equation} 
  M_{\tilde \nu \, LL \, ij}^2 
  =   m_{\tilde L \, ij}^2 + \left( 
      \frac{1}{2} \MZ^2 \cos 2\beta \right) \delta_{ij}.   
\label{eq:sneu-matrix}
\end{equation} 
 
It is important to note that due to $SU(2)_L$ gauge invariance
the same soft masses $m_{\tilde L \, ij}$ enter in
both the slepton and sneutrino $LL$ mass matrices. 
The soft SUSY-breaking parameters of the sneutrinos would differ from the corresponding ones
for charged sleptons by a rotation with the PMNS matrix. However, taking
the neutrino masses and oscillations 
into account in the SM leads to LFV effects that are extremely small. 
(For instance, in $\mu \to e \gamma$  they are of \order{10^{-47}} in case
of Dirac neutrinos with mass around 1~eV and maximal
mixing~\cite{Kuno:1999jp,DiracNu,MajoranaNu}, and of \order{10^{-40}} in case
of Majorana neutrinos~\cite{Kuno:1999jp,MajoranaNu}.) Consequently we do not
expect large effects from the inclusion of neutrino mass effects here
and neglect a rotation with the PMNS matrix. 
The slepton mass matrix in terms of the $\deFABij$ is given as

\noindent \begin{equation}  
m^2_{\tilde L}= \left(\begin{array}{ccc}
 m^2_{\tilde L_{1}} & \delta_{12}^{LLL} m_{\tilde L_{1}}m_{\tilde L_{2}} & \delta_{13}^{LLL} m_{\tilde L_{1}}m_{\tilde L_{3}} \\
 \delta_{21}^{LLL} m_{\tilde L_{2}}m_{\tilde L_{1}} & m^2_{\tilde L_{2}}  & \delta_{23}^{LLL} m_{\tilde L_{2}}m_{\tilde L_{3}}\\
\delta_{31}^{LLL} m_{\tilde L_{3}}m_{\tilde L_{1}} & \delta_{32}^{LLL} m_{\tilde L_{3}}m_{\tilde L_{2}}& m^2_{\tilde L_{3}} \end{array}\right),\end{equation}

\noindent \begin{equation}
v_1 {\cal A}^l  =\left(\begin{array}{ccc}
m_e A_e & \delta_{12}^{ELR} m_{\tilde L_{1}}m_{\tilde E_{2}} & \delta_{13}^{ELR} m_{\tilde L_{1}}m_{\tilde E_{3}}\\
\delta_{21}^{ELR}  m_{\tilde L_{2}}m_{\tilde E_{1}} & m_\mu A_\mu & \delta_{23}^{ELR} m_{\tilde L_{2}}m_{\tilde E_{3}}\\
\delta_{31}^{ELR}  m_{\tilde L_{3}}m_{\tilde E_{1}} & \delta_{32}^{ELR}  m_{\tilde L_{3}} m_{\tilde E_{2}}& m_{\tau}A_{\tau}\end{array}\right),\label{v1Al}\end{equation}

\noindent \begin{equation}  
m^2_{\tilde E}= \left(\begin{array}{ccc}
 m^2_{\tilde E_{1}} & \delta_{12}^{ERR} m_{\tilde E_{1}}m_{\tilde E_{2}} & \delta_{13}^{ERR} m_{\tilde E_{1}}m_{\tilde E_{3}}\\
 \delta_{21}^{ERR} m_{\tilde E_{2}}m_{\tilde E_{1}} & m^2_{\tilde E_{2}}  & \delta_{23}^{ERR} m_{\tilde E_{2}}m_{\tilde E_{3}}\\
\delta_{31}^{ERR}  m_{\tilde E_{3}} m_{\tilde E_{1}}& \delta_{32}^{ERR} m_{\tilde E_{3}}m_{\tilde E_{2}}& m^2_{\tilde E_{3}} \end{array}\right).\end{equation}

We need to rotate the sleptons and sneutrinos from the electroweak interaction basis to the physical mass eigenstate basis, 
\BE
\VL  \til_{1} \\ \til_{2}  \\ \til_{3} \\
                                    \til_{4}   \\ \til_{5}  \\\til_{6}   \VR
  \; = \; R^{\til}  \VL \SelL \\ \SmuL \\\StauL \\ 
  \SelR \\ \SmuR \\ \StauR \VR ~,~~~~
\VL  \tinu_{1} \\ \tinu_{2}  \\  \tinu_{3}  \VR             \; = \; R^{\tinu}  \VL \tinu_{eL} \\ \tinu_{\mu L}  \\  \tinu_{\tau L}   \VR ~,
\label{rotsquarks}
\end{equation} 
with $R^{\til}$ and $R^{\tinu}$ being the respective $6\times 6$ and
$3\times 3$ unitary rotating matrices that yield the diagonal
mass-squared matrices as follows, 
\BEA
{\rm diag}\{m_{\til_1}^2, m_{\til_2}^2, 
          m_{\til_3}^2, m_{\til_4}^2, m_{\til_5}^2, m_{\til_6}^2 
           \}  & = &
R^{\til}  \;  {\cal M}_{\til}^2   \; 
 R^{\til \dagger}    ~,\\
{\rm diag}\{m_{\tinu_1}^2, m_{\tinu_2}^2, 
          m_{\tinu_3}^2  
          \}  & = &
R^{\tinu}  \;   {\cal M}_{\tinu}^2   \; 
 R^{\tinu \dagger}    ~.
 \EEA


%% file: sec3_observables.tex
\section{Observation of SUSY LFV at the EW scale}

\begin{figure}
\begin{center}
\epsfig{file=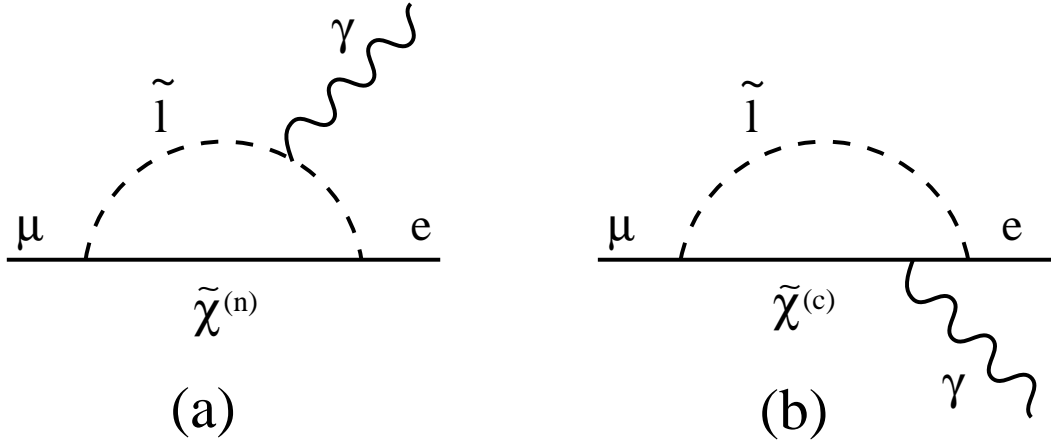,width=14cm}
\end{center}
\caption{The generic Feynman diagrams for the $\mu\rightarrow e\gamma$
decay. $\tilde l$ stands for charged slepton (a) or sneutrino (b), while
$\tilde\chi ^{(n)}$ and $\tilde\chi ^{(c)}$ represent  
neutralinos and charginos respectively.}
\label{figure1}
\end{figure}

SUSY particles enter in  SM processes at the loop level. Therfore, there is a SUSY contribution to processes predicted in the SM like the $b\rightarrow  s \gamma$. However, the equivalent  cLFV decays would arise only from loops medated by SUSY particles as the one of \reffi{figure1}. The bounds from the experimental search for these processes can be used to impose limits on the  $\deFABij$.  The aim of this paper is to evaluate the impact 
of the allowed $\deFABij$ on LFV Higgs decays. In this section we will review the observables that will be studied in the consecutive sections.

\subsection{Charged lepton flavor violating decays}

 Radiative LFV decays, $\mu\to e\gamma$, $\tau\to e\gamma$, and
$\tau\to \mu\gamma$ are sensitive to the $\deFABij$'s via the
$(\ell_i \ell_j\gamma)_{\text{1-loop}}$ vertices with a real photon. Fig.\ref{figure1} shows the one-loop diagrams relevant to the $\mu\ra
e\gamma$ process. The corresponding $\tau\ra \mu\gamma$ decay is
represented by an analogous set of graphs.

The electromagnetic current operator between two lepton  states $l_i$
and $l_j$ is given in general by
\begin{eqnarray}
{\cal T}_\la &=& \langle l_i(p-q)|{\cal J}_\la|l_j(p)\rangle\nonumber\\
{  }&=&{\bar u_i}(p-q)
      \{ m_j i\sigma_{\la\beta}q^\beta 
               \left(A^L_MP_L+A^R_MP_R\right)
      \} u_j(p)
\label{general}
\end{eqnarray}
where $q$ is the photon momentum. The $A_M$'s  receive
contributions from neutralino-charged slepton ($n$) and
chargino-sneutrino ($c$) exchange
\begin{equation}
A_M^{L,R}=A_{M(n)}^{L,R}+A_{M(c)}^{L,R}
\label{ampl}
\end{equation}
The Branching Ratio  of the decay $l_j\ra l_i+\gamma$ is given by
\[
BR(l_j\ra l_i\gamma)=\frac{48\pi^3\alpha}{G_F^2}
               \left((A_M^L)^2+(A_M^R)^2\right).
\]
\label{sec:lfv}

The above set of decay processes gives the most restrictive constraints on the  slepton $\deFABij$. Other cLFV decays which are sensitive to $\deFABij$ are also possible \cite{Arana-Catania:2013nha} :  

\begin{enumerate}
\item Leptonic LFV decays: $\mu\to 3 e$, $\tau\to 3 e$, and $\tau\to 3 
\mu$.  These are sensitive to the $\deFABij$'s via the 
$(\ell_i\ell_j\gamma)_{\text{1-loop}}$ vertices with a virtual photon, 
via the $(\ell_i\ell_j Z)_{\text{1-loop}}$ vertices with a virtual $Z$, 
and via the $(\ell_i\ell_j h)_{\text{1-loop}}$, $(\ell_i\ell_j 
H)_{\text{1-loop}}$ and $(\ell_i\ell_j A)_{\text{1-loop}}$ vertices with 
virtual Higgs bosons.

\item Semileptonic LFV tau decays: $\tau\to \mu\eta$ and $\tau\to 
e\eta$.  These are sensitive to the $\deFABij$'s via the $(\tau\ell 
A)_{\text{1-loop}}$ vertex with a virtual $A$ and the $(\tau\ell 
Z)_{\text{1-loop}}$ vertex with a virtual $Z$, where $\ell = \mu, e$, 
respectively.

\item Conversion of $\mu$ into $e$ in heavy nuclei: These are sensitive 
to the $\deFABij$'s via the $(\mu e\gamma)_{\text{1-loop}}$ vertex with a 
virtual photon, the $(\mu e Z)_{\text{1-loop}}$ vertex with a virtual 
$Z$, and the $(\mu e h)_{\text{1-loop}}$ and $(\mu e H)_{\text{1-loop}}$ 
vertices with a virtual Higgs boson.
\end{enumerate}

However, the indirect bounds that can be obtained on the lepton flavor violating $\deFABij$'s from these processes are less restrictive than the ones from radiative LFV decays. Present experimental limits
on these decay processes are summerized in \refta{cLFV:Exp-Limits}:

\begin{table}[h!]
\begin{center}
\begin{tabular}{|c|c||c|c|}
\hline
 observable &  experimental limit &  observable &  experimental limit  \\ 
\hline
$\br(\mu \to e \gamma)$ & $5.7 \times 10^{-13}$ & $\br(\tau \to e e e)$ &   $2.7 \times 10^{-8}$ \\ 
\hline
$\br(\tau \to \mu \gamma)$ & $ 4.4 \times 10^{-8}$ & $ \CR(\mu-e, {\rm Au}) $ & $7.0 \times 10^{-13}$ \\ 
\hline
$\br(\tau \to e \gamma)$ & $3.3 \times 10^{-8}$& $\br(\tau \to \mu \eta)$ & $2.3\times10^{-8}$ \\ 
\hline
$ \br(\mu \to eee) $ &  $1.0 \times 10^{-12}$ & $\br(\tau \to e \eta)$ & $4.4\times10^{-8}$ \\ 
\hline
 $\br(\tau \to \mu\mu\mu) $ & $2.1 \times 10^{-8}$  &   &  \\ 
 \hline
\end{tabular}
\caption{Present experimental limits on the cLFV decays\cite{Adam:2013mnn,Aubert:2009ag,Bellgardt:1987du,Bertl:2006up,Hayasaka:2010np}.}
\label{cLFV:Exp-Limits}
\end{center}
\end{table}

\subsection{Lepton flavor violating Higgs decays}
\label{sec:LFVHD}


Since the discovery of a Higgs boson, special effort has been made to determine its properties. The motivation
for such an effort resides on understanding the mechanism for electroweak symmetry breaking. At present, several
aspects of the Higgs boson are to some extent well known, in particular those related with some of its expected
“standard” decay modes, namely: $WW^*$, $ZZ^*$, $\gamma \gamma$, $b \bar{b}$ and $\tau^+ \tau^-$ \cite{Khachatryan:2016vau}. Currently, measurements of these decay modes have
shown compatibility with the SM expectations, although with large associated uncertainties \cite{deFlorian:2016spz}.
Indeed, it is due to these large uncertainties that there is still room for non-standard decay properties, something
that has encouraged such searches at the LHC as well. Searches for invisible Higgs decays have been published in
\cite{CMSAad2014,CMSChatrchyan2014}. The CMS collaboration using the 2012 dataset taken at $\sqrt{s} = 8\; \rm TeV$ with an integrated luminosity of 19.7 $\rm fb^{-1}$, has found a 2.4 $\sigma$ excess in the $h\rightarrow \mu \tau$ channel, which translates into BR$(h\rightarrow \mu \tau) \approx 0.84^{+0.39}_{-0.37} \%$ \cite{cms}\footnote{The CMS collaboration released a new result\cite{CMS_not}, not published yet,  using  data taken  at $\sqrt{s} = 13\; \rm TeV$ corresponds to an integrated luminosity of 2.3 fb−1. No excess is observed at  95\% CL }. That is consistent with the less statistically significant excess, BR$(h\rightarrow \mu \tau)=(0.53 \pm 0.51) \%$,  reported by  ATLAS \cite{atlas}.

\begin{figure}[htb!]
\begin{center}
\vspace{0.2cm}
\psfig{file=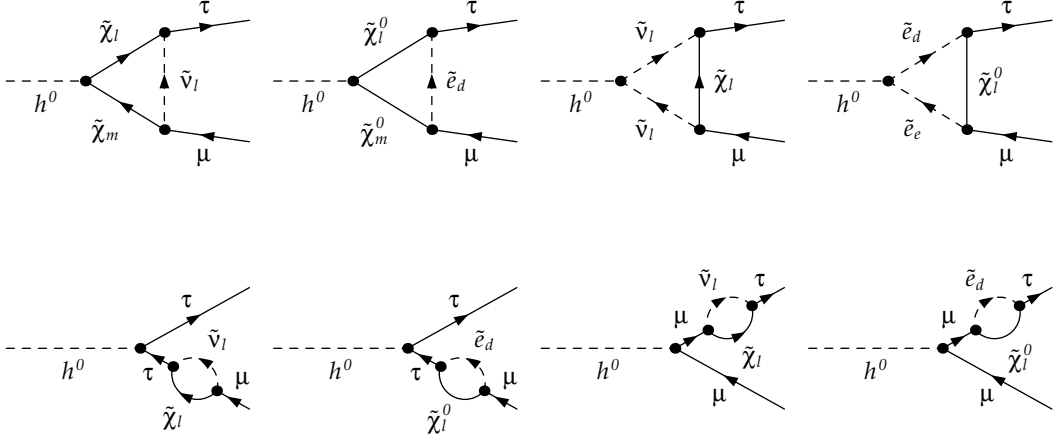}
\end{center}
\caption{Feynman diagrams for LFV decays $h \rightarrow \mu^{\pm} \tau^{\mp}$.}
\label{fig:Diag_LFVHD}
\end{figure}

Feynman diagrams for the process $h\rightarrow \mu \tau$ are dispalyed in \reffi{fig:Diag_LFVHD}. Using our \fa\ and \fc\ setup we can compute the branching ratios for the Higgs LFV decays in the context of the models under 
consideration. For numerical analysis we define the branching ratios of LFVHD as
\BE
BR(h \rightarrow l_i^{\pm} l_j^{\mp})= \frac{\Gamma(h \rightarrow l_i^{\pm} l_j^{\mp})}{\Gamma(h \rightarrow l_i^{\pm} l_j^{\mp})+\Gamma_h^{\rm MSSM}} 
\EE 
Where $i,j=e, \mu, \tau$ and $\Gamma_h^{\rm MSSM}$ is total decay width of $\cp$-even light Higgs boson without flavor violation.

%% file: sec4_MI_analysis.tex

\section{Model independent analysis}
\label{sec:NR_modelInd}
In this section we choose a model independent approach to perform the numerical analysis. As a framework we choose some MSSM model points compatible with present data, including recent LHC searches and the measurements of the muon anomalous magnetic moment. In addition, we include the range of values of $|\delta^{FAB}_{ij}|$ allowed from the current bounds on LFV decays.

\subsection{Input Parameters} 

For the following numerical analysis we chose the MSSM parameter sets of Refs.~\cite{drhoLFV, hbs2016, Arana-Catania:2013nha}.  The values of the 
various MSSM parameters as well as the values of the predicted MSSM mass 
spectra are summarized in \refta{tab:spectra}.  They were evaluated with 
the program 
\fh~\cite{feynhiggs,mhiggslong,mhiggsAEC,mhcMSSMlong,Mh-logresum}.

For simplicity, and to reduce the number of independent MSSM input 
parameters, we assume equal soft masses for the sleptons of the first 
and second generations (similarly for the squarks), and for the left and 
right slepton sectors (similarly for the squarks).  We choose equal 
trilinear couplings for the stops and sbottoms and for the 
sleptons consider only the stau trilinear coupling; the others are set 
to zero.  We assume an approximate GUT relation for the gaugino 
soft-SUSY-breaking parameters.  The pseudoscalar Higgs mass $\MA$ and 
the $\mu$ parameter are taken as independent input parameters.  In 
summary, the six points S1\dots S6 are defined in terms of the 
following subset of ten input MSSM parameters at the SUSY scale:
\begin{align*}
&m_{\tilde L_1} = m_{\tilde L_2}\,, &
&m_{\tilde L_3}\,, &
&(\text{with~} m_{\tilde L_i} = m_{\tilde E_i},\ i = 1,2,3) \\
&m_{\tilde Q_1} = m_{\tilde Q_2} &
&m_{\tilde Q_3}\,, &
&(\text{with~} m_{\tilde Q_i} = m_{\tilde U_i} = m_{\tilde D_i},\ i = 1,2,3) \\
&A_t = A_b\,, &
&A_\tau\,, \\
&M_2 = 2 M_1 = M_3/4\,, &
&\mu\,, \\
&\MA\,, &
&\tb\,.
\end{align*}

\begin{table}[h!]
\caption{\label{tab:spectra}Selected points in the MSSM parameter space 
(upper part) and their corresponding spectra  in the case of setting all the $\delta$'s to zero (lower part).  All 
dimensionful quantities are in GeV.}
\centerline{\begin{tabular}{|c|c|c|c|c|c|c|}
\hline
 & S1 & S2 & S3 & S4 & S5 & S6 \\\hline
$m_{\tilde L_{1,2}}$& 500 & 750 & 1000 & 800 & 500 &  1500 \\
$m_{\tilde L_{3}}$ & 500 & 750 & 1000 & 500 & 500 &  1500 \\
$M_2$ & 500 & 500 & 500 & 500 & 750 &  300 \\
$A_\tau$ & 500 & 750 & 1000 & 500 & 0 & 1500  \\
$\mu$ & 400 & 400 & 400 & 400 & 800 &  300 \\
$\tb$ & 20 & 30 & 50 & 40 & 10 & 40  \\
$\MA$ & 500 & 1000 & 1000 & 1000 & 1000 & 1500  \\
$m_{\tilde Q_{1,2}}$ & 2000 & 2000 & 2000 & 2000 & 2500 & 1500  \\
$m_{\tilde Q_{3}}$ & 2000 & 2000 & 2000 & 500 & 2500 & 1500  \\
$A_t$ & 2300 & 2300 & 2300 & 1000 & 2500 &  1500 \\\hline
$m_{\til_{1\dots 6}}$ & 489--515 & 738--765 & 984--1018 & 474--802  & 488--516 & 1494--1507  \\
$m_{\tinu_{1\dots 3}}$ & 496 & 747 & 998 & 496--797 & 496 &  1499 \\
$m_{{\tilde\chi}_{1,2}^\pm}$ & 375--531 & 376--530 & 377--530 & 377--530  & 710--844 & 247--363  \\
$m_{{\tilde\chi}^0_{1\dots 4}}$ & 244--531 & 245--531 & 245--530 & 245--530  & 373--844 & 145--363  \\
$M_h$ & 126.6 & 127.0 & 127.3 & 123.1 & 123.8 & 125.1  \\
$M_H$ & 500 & 1000 & 999 & 1001 & 1000 & 1499  \\
$M_A$ & 500 & 1000 & 1000 & 1000 & 1000 & 1500  \\
$M_{H^\pm}$ & 507 & 1003 & 1003 & 1005 & 1003 & 1502  \\
$m_{\tilde u_{1\dots 6}}$& 1909--2100 & 1909--2100 & 1908--2100 & 336--2000 & 2423--2585 & 1423--1589  \\
$m_{\tilde d_{1\dots 6}}$ & 1997--2004 & 1994--2007 & 1990--2011 & 474--2001 & 2498--2503 &  1492--1509 \\
$m_{\tilde g}$ & 2000 & 2000 & 2000 & 2000 & 3000 &  1200 \\
\hline
\end{tabular}}
\end{table}

The specific values of these ten MSSM parameters in \refta{tab:spectra} 
are chosen to provide different patterns in the various sparticle 
masses, but all leading to rather heavy spectra and thus naturally in 
agreement with the absence of SUSY signals at the LHC.  In particular, 
all points lead to rather heavy squarks and gluinos above $1200\gev$ and 
heavy sleptons above $500\gev$ (where the LHC limits would also permit 
substantially lighter sleptons).  The values of $\MA$ within the 
interval $(500,1500)\gev$, $\tb$ within the interval $(10,50)$ and a 
large $A_t$ within $(1000,2500)\gev$ are fixed such that a light Higgs 
boson $h$ within the LHC-favoured range $(123,127)\gev$ is obtained.

The large values of $\MA\geqslant 500$ GeV place the Higgs sector of our 
scenarios in the so-called decoupling regime\cite{Haber:1989xc}, where 
the couplings of $h$ to gauge bosons and fermions are close to the SM 
Higgs couplings, and the heavy $H$ couples like the pseudoscalar $A$, 
and all heavy Higgs bosons are close in mass.  With increasing $\MA$, 
the heavy Higgs bosons tend to decouple from low-energy physics and the 
light $h$ behaves like the SM Higgs.  This type of MSSM Higgs sector 
seems to be in good agreement with recent LHC data\cite{LHCHiggs}.  We 
checked with the code HiggsBounds~\cite{higgsbounds} that this is indeed 
the case (although S3 is right `at the border').

Particularly, the absence of gluinos at the LHC so far forbids too low 
$M_3$ and, through the assumed GUT relation, also a too low $M_2$.  This 
is reflected by our choice of $M_2$ and $\mu$ which give gaugino masses 
compatible with present LHC bounds.  Finally, we required that all our 
points lead to a prediction of the anomalous magnetic moment of the muon 
in the MSSM that can fill the present discrepancy between the Standard 
Model prediction and the experimental value.

\renewcommand{\arraystretch}{1.1}
\begin{table}[htb!]
\begin{center}
\resizebox{15.0cm}{!} {
\begin{tabular}{|c|c|c|c|c|c|c|}
\hline
 &  S1 &  S2 &  S3 &  S4 &  S5 & S6  
 \\ \hline
 & & & & & & \\
$|\delta^{LLL}_{12}|_{\rm max}$ & $10 \times 10^{-5}$ & $7.5\times 10^{-5}$ &   $5 \times 10^{-5}$& $6 \times 10^{-5}$ & $42\times 10^{-5}$  &  $8\times 10^{-5}$  \\ 
& & & & & & \\
\hline
& & & & & & \\
$|\delta^{ELR}_{12}|_{\rm max}$ & $2\times 10^{-6}$ & $3\times 10^{-6}$ &
$4\times 10^{-6}$  & $3\times 10^{-6}$ & $2\times 10^{-6}$  & $1.2\times 10^{-5}$   \\ 
& & & & & & \\
\hline
& & & & & & \\
$|\delta^{ERR}_{12}|_{\rm max}$ & $1.5 \times 10^{-3}$& $1.2 \times 10^{-3}$ & 
$1.1 \times 10^{-3}$ & $1 \times 10^{-3}$ & $2 \times 10^{-3}$ & $5.2 \times 10^{-3}$   \\ 
& & & & & & \\
\hline
& & & & & & \\
$|\delta^{LLL}_{13}|_{\rm max} $ &  $5 \times 10^{-2}$ & $5 \times 10^{-2}$ & 
$3 \times 10^{-2}$ &  $3 \times 10^{-2}$& $23 \times 10^{-2}$ & $5 \times 10^{-2}$   \\ 
& & & & & & \\
\hline
& & & & & & \\
 $|\delta^{ELR}_{13}|_{\rm max}$& $2\times 10^{-2}$  & $3\times 10^{-2}$ & $4\times 10^{-2}$ & $2.5\times 10^{-2}$ & $2\times 10^{-2}$ & $11\times 10^{-2}$   \\ 
& & & & & & \\
 \hline
 & & & & & & \\
$|\delta^{ERR}_{13}|_{\rm max}$ & $5.4\times 10^{-1}$  & $5\times 10^{-1}$ & 
 $4.8\times 10^{-1}$ &$5.3\times 10^{-1}$  & $7.7\times 10^{-1}$ & $7.7\times 10^{-1}$ 
  \\ 
 & & & & & & \\ 
  \hline
 & & & & & & \\ 
$|\delta^{LLL}_{23}|_{\rm max}$ & $6\times 10^{-2}$  & $6\times 10^{-2}$ & 
 $4\times 10^{-2}$& $4\times 10^{-2}$ & $27\times 10^{-2}$ & $6\times 10^{-2}$ 
  \\ 
 & & & & & & \\ 
  \hline
 & & & & & & \\ 
$|\delta^{ELR}_{23}|_{\rm max}$ & $2\times 10^{-2}$   & $3\times 10^{-2}$ & 
$4\times 10^{-2}$ & $3\times 10^{-2}$ & $2\times 10^{-2}$ & $12\times 10^{-2}$ 
  \\ 
 & & & & & & \\ 
  \hline
 & & & & & & \\ 
$|\delta^{ERR}_{23}|_{\rm max}$ & $5.7\times 10^{-1}$  & $5.2\times 10^{-1}$ & 
 $5\times 10^{-1}$& $5.6\times 10^{-1}$ & $8.3\times 10^{-1}$ & $8\times 10^{-1}$ 
  \\ 
 & & & & & & \\  
  \hline
\end{tabular}}  
\end{center}
\caption[Constraints on $|\delta^{FAB}_{ij}|$ from LFV decays.]{ Present upper bounds on the slepton mixing parameters $|\delta^{FAB}_{ij}|$ for the selected S1-S6 MSSM points defined in \refta{tab:spectra}. The bounds for $|\delta^{ERL}_{ij}|$ are similar 
to those of $|\delta^{ELR}_{ij}|$.}
\label{boundsSpoints}
\end{table}


Applying the most recent limits from the above listed LFV process yield
up-to-date limits on the $\deFABij$~\cite{Arana-Catania:2013nha}. 
Using the these upper bounds on $\delta^{FAB}_{ij}$, as given in the
\refta{boundsSpoints}, we calculate the  predictions for LFV Higgs decays.

\subsection{{\boldmath ${\rm BR}(h \rightarrow l_i^{\pm} l_j^{\mp})$}}
\begin{figure}[ht!]
\begin{center}
\psfig{file=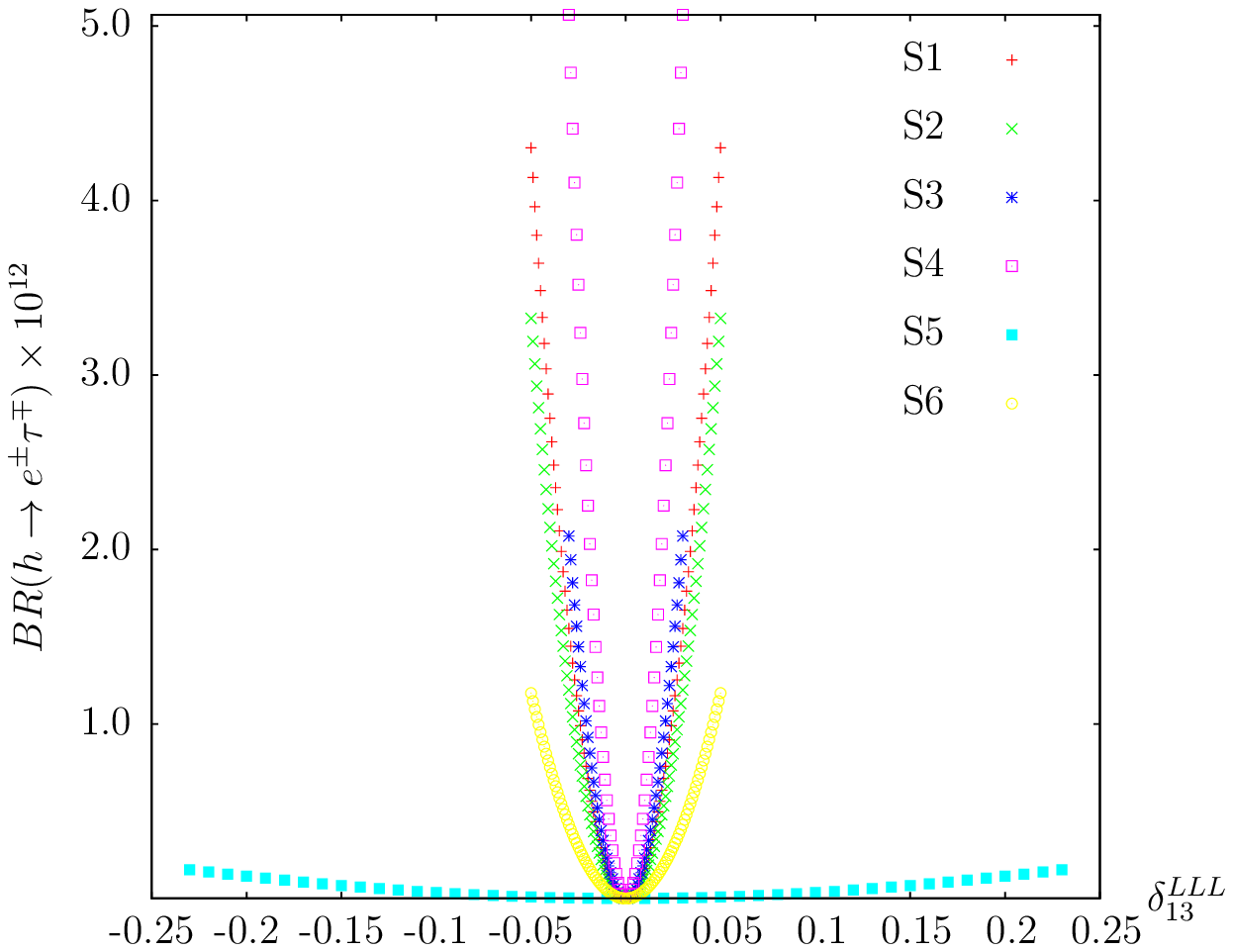  ,scale=0.52,angle=0,clip=}
\psfig{file=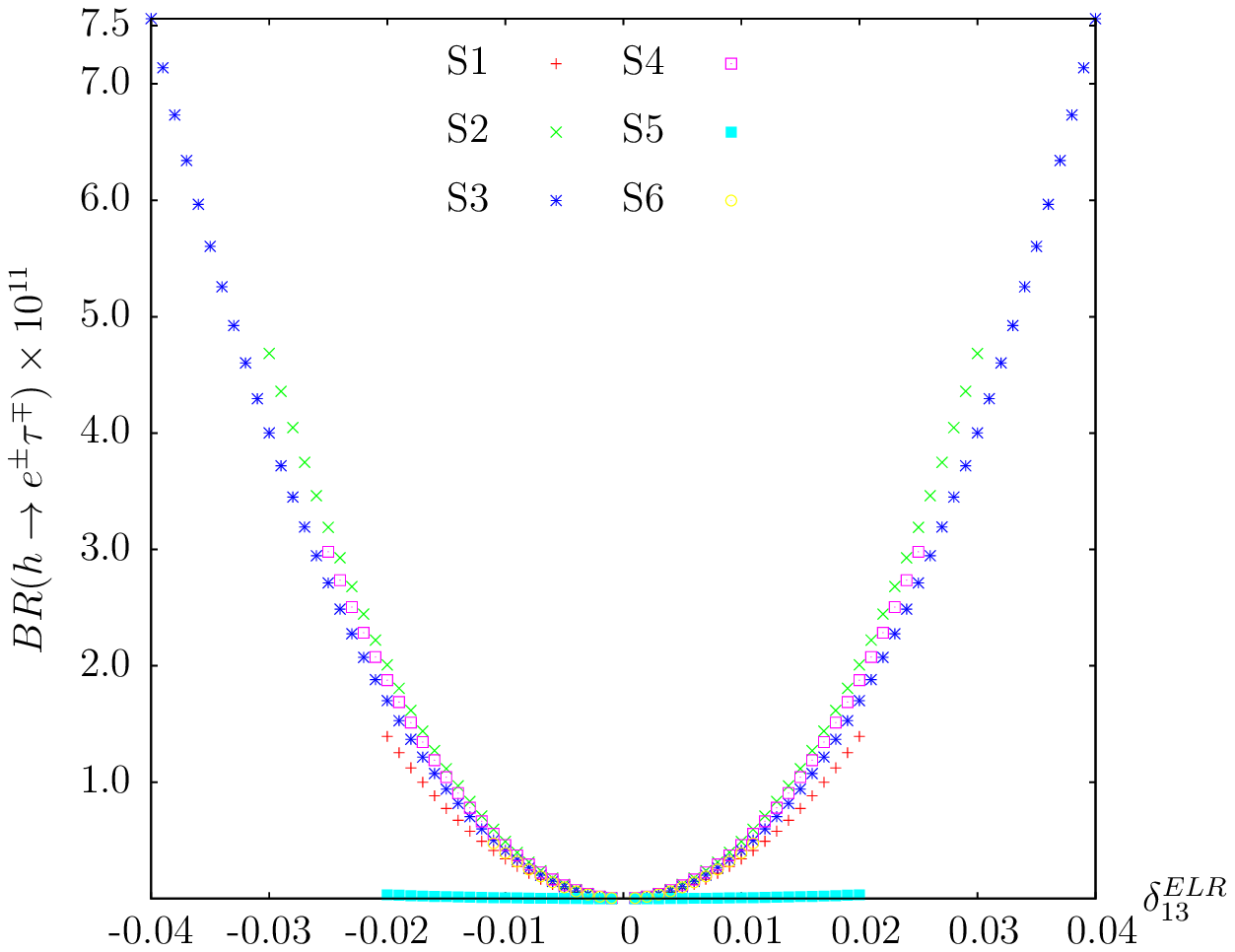  ,scale=0.52,angle=0,clip=}\\
\vspace{0.7cm}
\psfig{file=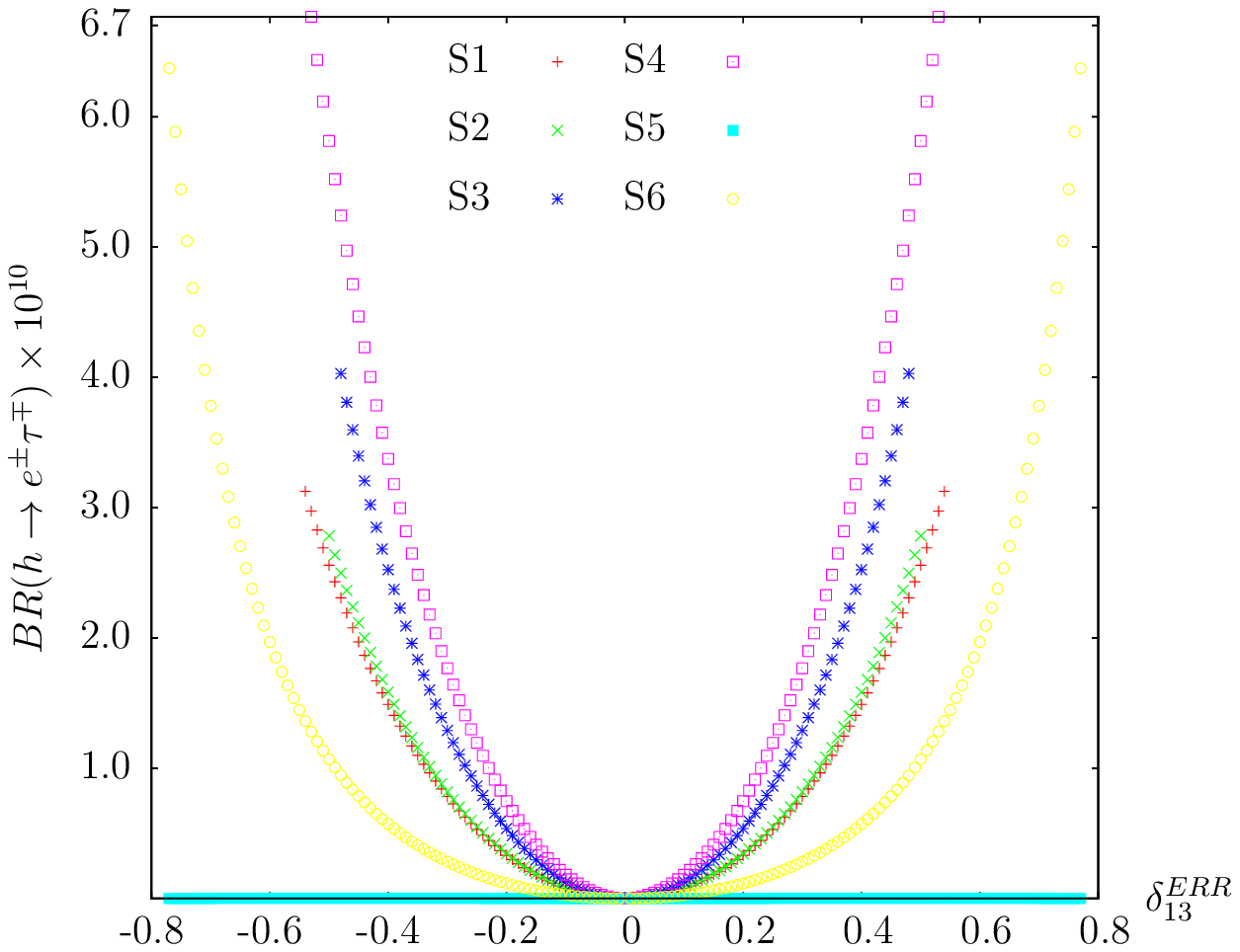 ,scale=0.52,angle=0,clip=}
\psfig{file=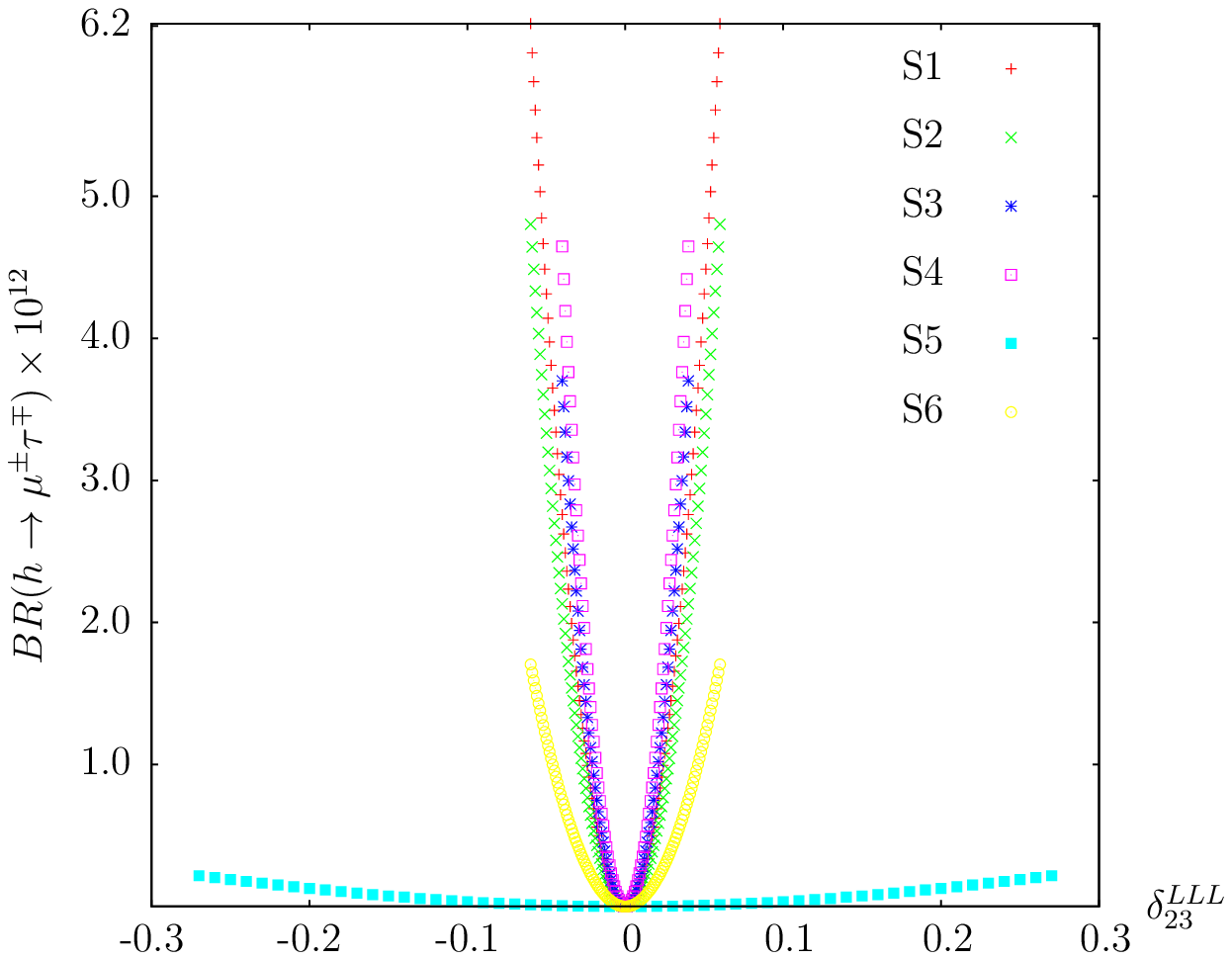   ,scale=0.52,angle=0,clip=}\\
\vspace{0.7cm}
\psfig{file=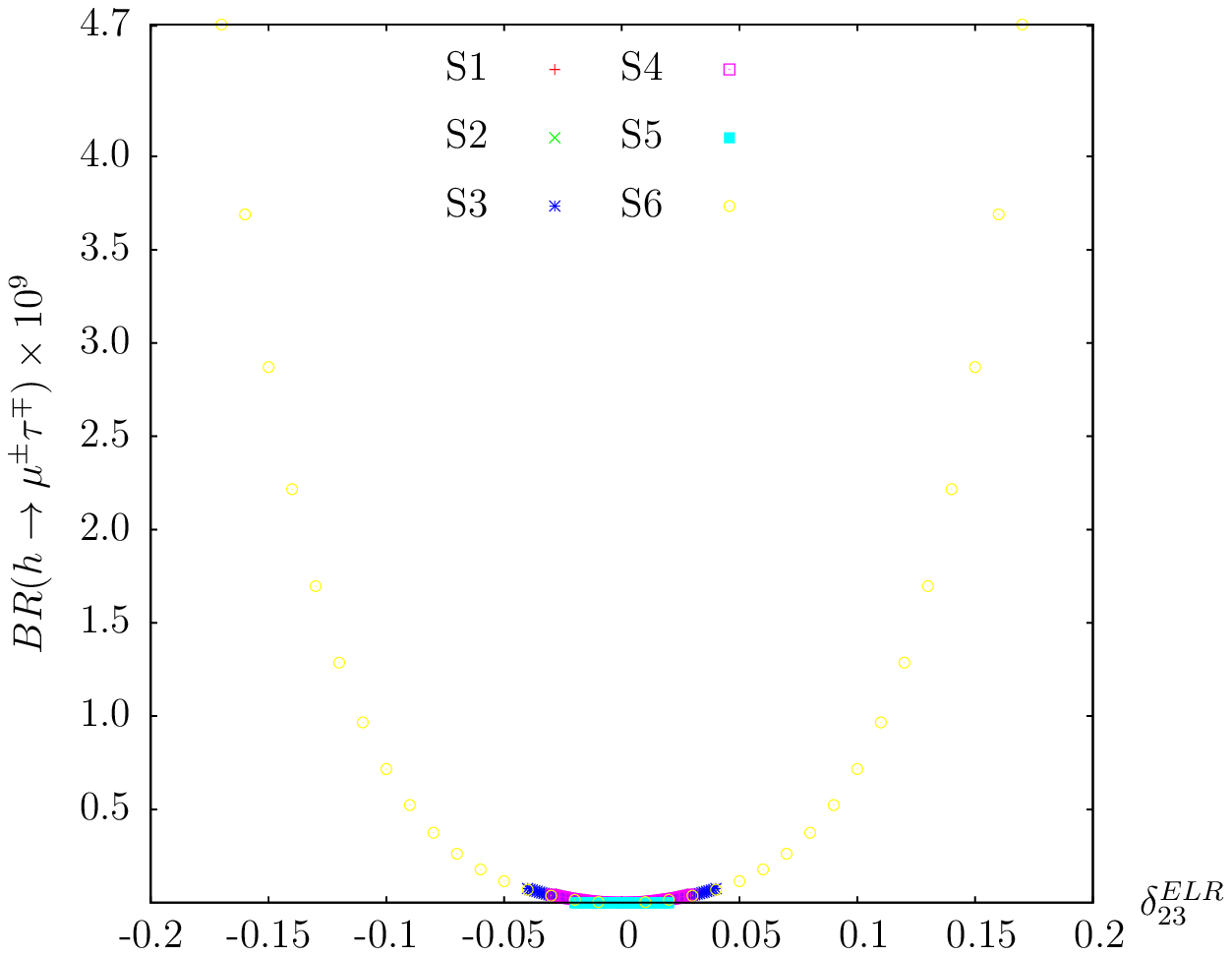  ,scale=0.52,angle=0,clip=}
\psfig{file=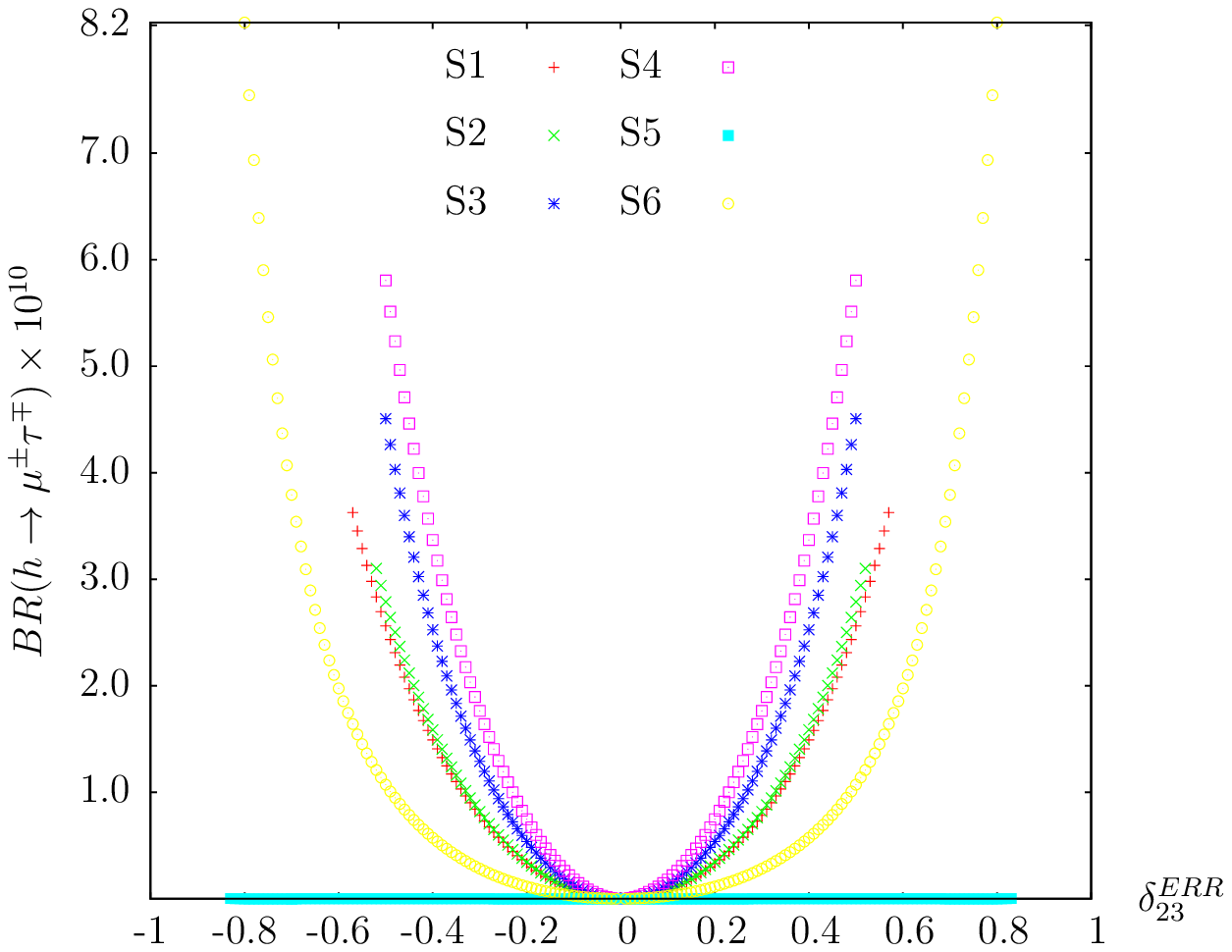  ,scale=0.52,angle=0,clip=}\\

\end{center}
\caption{Lepton flavor violating decays $h \rightarrow e \tau $ and $h \rightarrow \mu \tau $ as a function of slepton
  mixing $\deFABij$ for the six points defined in the \refta{tab:spectra}.}  
\label{fig:Hetau:Hmutau}
\end{figure}

We present here the slepton mixing effects to the LFVHD. These decays were calculated using newly modified \fa/\fc\ setup. 
The constraints from cLFV decays on slepton $\deFABij$'s are very tight and we do not expect large values for the BR's.  In \reffi{fig:Hetau:Hmutau} we present our numerical results for BR($h \rightarrow e^{\pm} \tau^{\mp} $) and BR($h \rightarrow \mu^{\pm} \tau^{\mp} $) as a function of slepton mixing $\deFABij$'s for the six points defined in the \refta{tab:spectra}. ${\rm BR}(h \rightarrow e^{\pm} \mu^{\mp})$ can only reach \order{10^{-17}} at maximum and we do not show them here. BR($h \rightarrow e^{\pm} \tau^{\mp} $) and BR($h \rightarrow \mu^{\pm} \tau^{\mp} $) can reach at most to \order{10^{-9}} for some parameter points, which is very small compared to an excess at the level of  the  original CMS excess \cite{cms}. Such small values are
expected because the diagrams shown in \reffi{fig:Diag_LFVHD} contain the same neutralino and chargino couplings that appear in the cLFV decays of \reffi{figure1} with very strong experimental bounds \cite{Arganda:2015uca,Aloni:2015wvn}.  LFV Higgs interaction are enhanced in the non decoupling regime ($M_A\gtrsim M_Z$)  \cite{Babu:2002et,Dedes:2002rh,Cannoni:2008bg,Hisano:2010es} leading to larger values for BR($h \rightarrow \mu^{\pm} \tau^{\mp} $), like the ones found in Refs. \cite{Brignole:2003iv,Brignole:2004ah,Kanemura:2004cn,Paradisi:2005tk} however such values for $M_A$ are excluded on the MSSM  by  the $H/A \rightarrow \tau \tau$ searches \cite{a-t-t}.  Therfore,  in the framework considered here, some other sources of LFV will be required to explain a CMS-type  result in the case that it is confirmed in the future run of the LHC. Lepton-slepton misalignment is not sufficient to explain this excess.
  
%% file: sec5_CMSSM-SeesawI.tex
\section{Lepton Flavor Mixing Effects in the \CMSSMI\ }
\label{NR-CMSSMI}
After presenting the MI analysis in the previous section, here we investigate the predictions of the MSSM complemented with a ''see-saw´´ mechanism to explain neutrino masses. In this framework,  values for  $\deFABij$ are radiatively generated even if  the soft terms are assumed universal at the GUT scale. 

One of the simpler implementations of the ''see-saw´´ mechanism on the MSSM is the  type-I seesaw mechanism~\cite{seesaw:I}. The superpotential for MSSM-Seesaw I can be written as

\begin{eqnarray}
\label{superpotentialSeesaw1}
W&=&W_{\rm MSSM}+ Y_{\nu}^{ij}\epsilon_{\alpha \beta} H_2^{\alpha} N_i^c L_j^{\beta}
+ \frac{1}{2} M_{N}^{ij} N_i^c N_j^c,
\end{eqnarray}
\noindent
where $W_{\rm MSSM}$ is given in \refeq{superpotential} and $N_i^c$
is the additional superfield that contains the three right-handed neutrinos,
$\nu_{Ri}$, and 
their scalar partners, $\tilde \nu_{Ri}$. $M_N^{ij}$ denotes the $3\times3$ Majorana mass matrix for heavy right handed neutrino.
The full set of soft SUSY-breaking terms is given by,
\begin{eqnarray}
\label{softbreakingSeesaw1}
-\cL_{\rm soft,SI} &=& - \cL_{\rm soft}
+(m_{\tilde \nu}^2)^i_j {\tilde \nu}_{Ri}^* {\tilde \nu}_{R}^j
+ (\frac{1}{2}B_{\nu}^{ij} M_{N}^{ij} {\tilde \nu}_{Ri}^* {\tilde \nu}_{Rj}^*
+A_{\nu}^{ij}h_2 {\tilde \nu}_{Ri}^* {\tilde l}_{Lj}+ {\rm h.c.})~,
\end{eqnarray}
with $\cL_{\rm soft}$ given by \refeq{softbreaking}, 
$(m_{\tilde \nu}^2)^i_j$,  $A_{\nu}^{ij}$ and $B_{\nu}^{ij}$
are the new soft breaking parameters.

By the seesaw mechanism three of the neutral fields acquire heavy masses and
decouple at high energy scale that we will denote as $M_N$, below this scale the effective theory
contains the MSSM plus an operator that provides masses to the neutrinos.

\begin{equation}
W=W_{\rm MSSM}+ \frac{1}{2}(Y_{\nu} L  H_2)^{T}  M_{N}^{-1} (Y_{\nu} L  H_2).
\end{equation}

As right handed neutrinos decouple at their respective mass scales, at low energy we have the same particle content and mass matrices as in the MSSM. This framework naturally explains neutrino oscillations in agreement with
experimental data~\cite{Neutrino-Osc}. At the electroweak scale an
effective Majorana mass matrix for light neutrinos, 
\begin{equation}
\label{meff}
m_{\rm eff}=-\frac{1}{2}v_u^2 Y_{\nu}\cdot M_{N}^{-1}\cdot Y^{ T}_{\nu}, 
\end{equation}
arises from Dirac neutrino Yukawa $Y_{\nu}$ (that can be assumed of
the same order as the charged-lepton and quark Yukawas), and heavy Majorana
masses $M_N$.  The smallness of the neutrino masses implies that the scale
$M_N$ is very high, \order{10^{14} \gev}. 

From \refeqs{superpotentialSeesaw1} and (\ref{softbreakingSeesaw1}) 
we can observe that 
one can choose a basis such that the Yukawa coupling matrix,
$Y_l^{ij}$, and the mass matrix of the right-handed neutrinos, $M_N^{ij}$, are
diagonalized as $Y_l^\delta$ and $M_R^\delta$, respectively. In this case
the neutrino Yukawa couplings $Y_{\nu}^{ij}$ are not generally diagonal,
giving rise to LFV ~\cite{Cannoni:2013gq,gllv,Mismatch,Antusch,EGL,casas-ibarra} . Here it is important to note that the lepton-flavor
conservation is not a consequence of the SM gauge symmetry, even in the absence
of the right-handed neutrinos. 
Consequently, slepton mass terms can violate
the lepton-flavor conservation in a manner consistent with the gauge
symmetry.  Thus the scale of LFV can be identified with the
EW scale, much lower than the right-handed neutrino scale $M_N$. In
the basis where the charged-lepton masses $Y_{\ell}$ is diagonal, the soft
slepton-mass matrix acquires corrections that contain off-diagonal
contributions from the RGE running from $M_{\rm GUT}$ down to the 
Majorana mass scale $M_N$,
of the following form (in the leading-log approximation, assuming that $M_N$ is a common scale for the three heavy neutrino masses)~\cite{LFVhisano}: 
\begin{align}
(m_{\tilde L}^2)_{ij} &\sim \frac 1{16\pi^2} (6m^2_0 + 2A^2_0)
\left({Y_{\nu}}^{\dagger} Y_{\nu}\right)_{ij}  
\log \KL \frac{M_{\rm GUT}}{M_N} \KR \, \nonumber\\
(m_{\tilde e}^2)_{ij} &\sim 0  \, \nonumber\\
(A_l)_{ij} &\sim  \frac 3{8\pi^2} {A_0 Y_{l}}_i
\left({Y_{\nu}}^{\dagger} Y_{\nu}\right)_{ij}  
\log \KL \frac{M_{\rm GUT}}{M_N} \KR \,
\label{offdiagonal}
\end{align}
 Below this  
scale, the off-diagonal contributions remain almost unchanged.

The values of $\deFABij$ depend on the structure of $Y_\nu$ at a see-saw
scale $M_N$ in a basis where $Y_l$ and $M_N$ are diagonal.  
By using the approach of \citere{casas-ibarra} a general form of $Y_\nu$
containing all neutrino experimental information can be wtritten as: 

\begin{equation}
Y_\nu = \frac{\sqrt{2}} {v_u} \sqrt{M_R^\delta} R  \sqrt{m_\nu^\delta} U^\dagger~,
\label{eq:casas} 
\end{equation}
where $R$ is a general orthogonal matrix and $m_\nu^\de$ denotes the
diagonalized neutrino mass matrix. In this basis the matrix~$U$ can be
identified with the $U_{\rm PMNS}$ matrix obtained as: 
\begin{equation}
m_\nu^\delta=U^T m_{\rm eff} U~.
\end{equation}

In order to find values for the slepton generation mixing
parameters we need a
specific form of the product $Y_\nu^\dagger Y_\nu$ as shown in
\refeq{offdiagonal}. The 
simple consideration of direct hierarchical neutrinos with a common
scale for right handed neutrinos provides a representative reference
value. In this case using \refeq{eq:casas} we find 
\begin{equation}
Y_\nu^\dagger Y_\nu= \frac{2}{v_u^2}M_R U m_\nu^\delta U^\dagger~.
\label{eq:ynu2}
\end{equation}
Here $M_R$ is the common mass assigned to the $\nu_R$'s. In the conditions
considered here, LFV effects are independent of the matrix $R$.

In order to perform our calculations, we used {\tt SPheno}~\cite{Porod:2003um}
to generate the \CMSSMI\ particle
spectrum by running RGE from the GUT down to the EW scale. 
The particle spectrum was handed over in the form of an SLHA
file~\cite{SLHA} to 
\fa/\fc\ setup via \fh~\cite{feynhiggs,mhiggslong,mhiggsAEC,mhcMSSMlong,Mh-logresum} to
calculate  LFVHD whereas cLFV decays were calculated with {\tt SPheno 3.2.4}. The following section describes the details of our computational setup.
\subsection{Input Parameters}
\label{sec:GUTEW}

For our scans of the \CMSSMI\ parameter space we use 
{\tt SPheno 3.2.4}~\cite{Porod:2003um} with
the model ``see-saw type-I'' as in Ref.~\cite{Gomez:2015ila}. For the numerical analysis the values of the Yukawa couplings
etc.\ have to be set to yield values in agreement with the
experimental data for neutrino masses and mixings.
In our computation, by considering a normal  hierarchy among the neutrino 
masses, we fix 
$m_{\nu_3} \sim \sqrt{\Delta m^2_{\text{atm}}} \sim 0.05 \ev$ and
require $m_{\nu_2}/m_{\nu_3}=0.17$, 
$m_{\nu_2} \sim  100 \cdot m_{\nu_1}$ consistent with the measured values of 
$\Delta m^2_{\text{sol}}$ and $\Delta m^2_{\text{atm}}$~\cite{Neutrino-Osc}. 
The matrix $U$ in Eq.~\ref{eq:casas} is identified with $U_{\rm PMNS}$ with the $\cp$-phases set to
zero and neutrino mixing angles set to the center of their
experimental values.  When the $Y_\nu$ of Eq.~\ref{eq:casas} is constructed using these values for $m_\nu^\delta$ and common values for $M_R^\delta=M_N$ we find representative values for the $\deFABij$'s. Since these depend only on the product  $Y_\nu^\dagger Y_\nu$, they are independent on the orthogonal matrix $R$ that can be set equal to the identity.  
By setting $M_N=10^{14} \gev$, the values $Y_\nu$ remain perturbative.  An  
example of models  with almost degenerate $\nu_R$ can be found in
\cite{Cannoni:2013gq}. For our numerical analysis we tested several scenarios
and we found that the one defined here is the simplest and also the
one with larger LFV prediction.


In order to get an overview about the size of the effects in the \CMSSMI\
parameter space, the relevant parameters $m_0$, $m_{1/2}$ have been
scanned as, or in case of $A_0$ and
$\tb$ have been set to all combinations of 
\begin{align}
m_0 &\eq 500 \gev \ldots 5000 \gev~, \\
m_{1/2} &\eq 1000 \gev \ldots 3000 \gev~, \\
A_0 &\eq -3000, -2000, -1000, 0 \gev~, \\
\tb &\eq 10, 20, 35, 45~,
\end{align}
with $\mu > 0$.  

\begin{figure}[ht!]
\begin{center}
\psfig{file=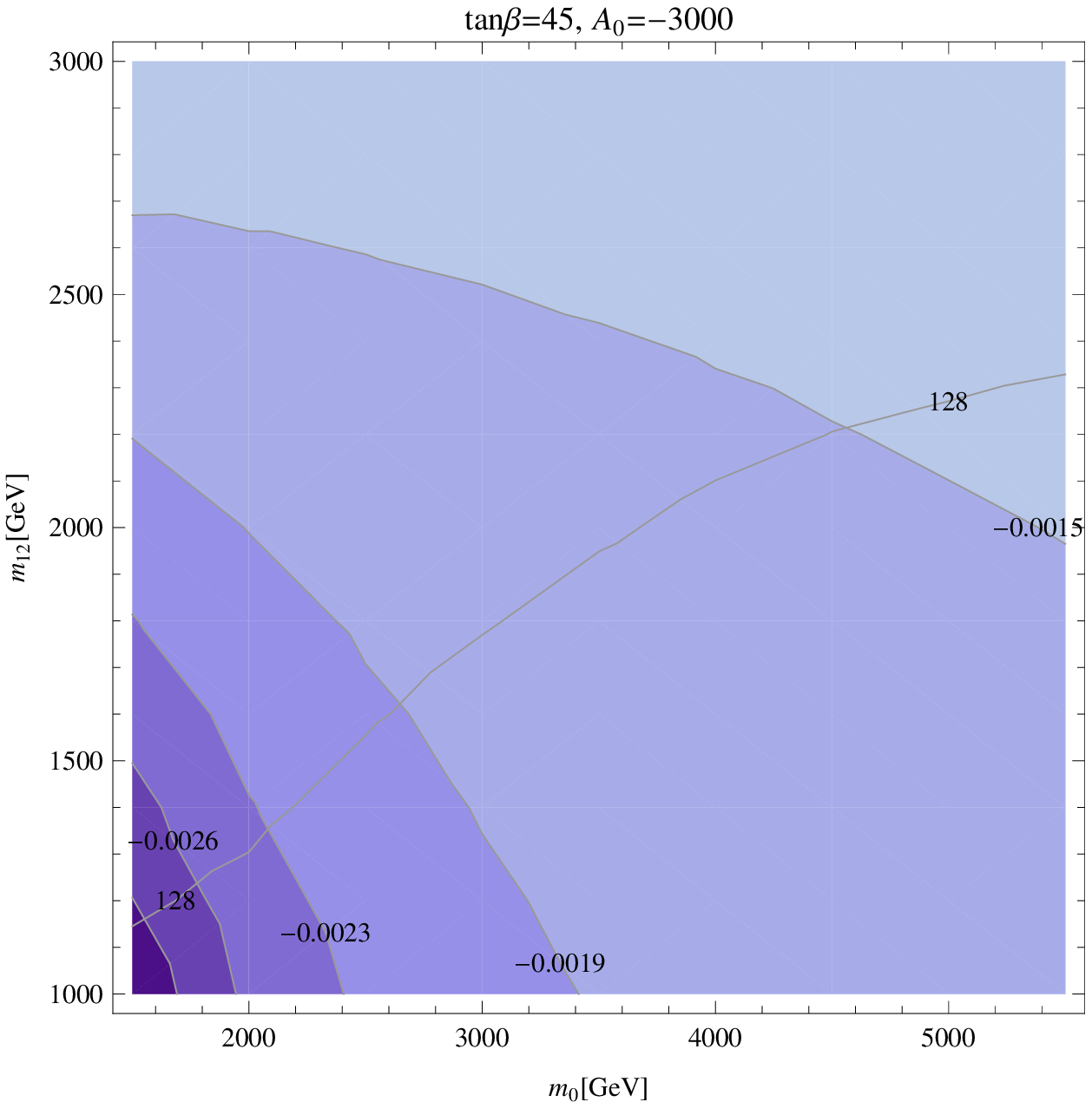 ,scale=0.51,angle=0,clip=}
\psfig{file=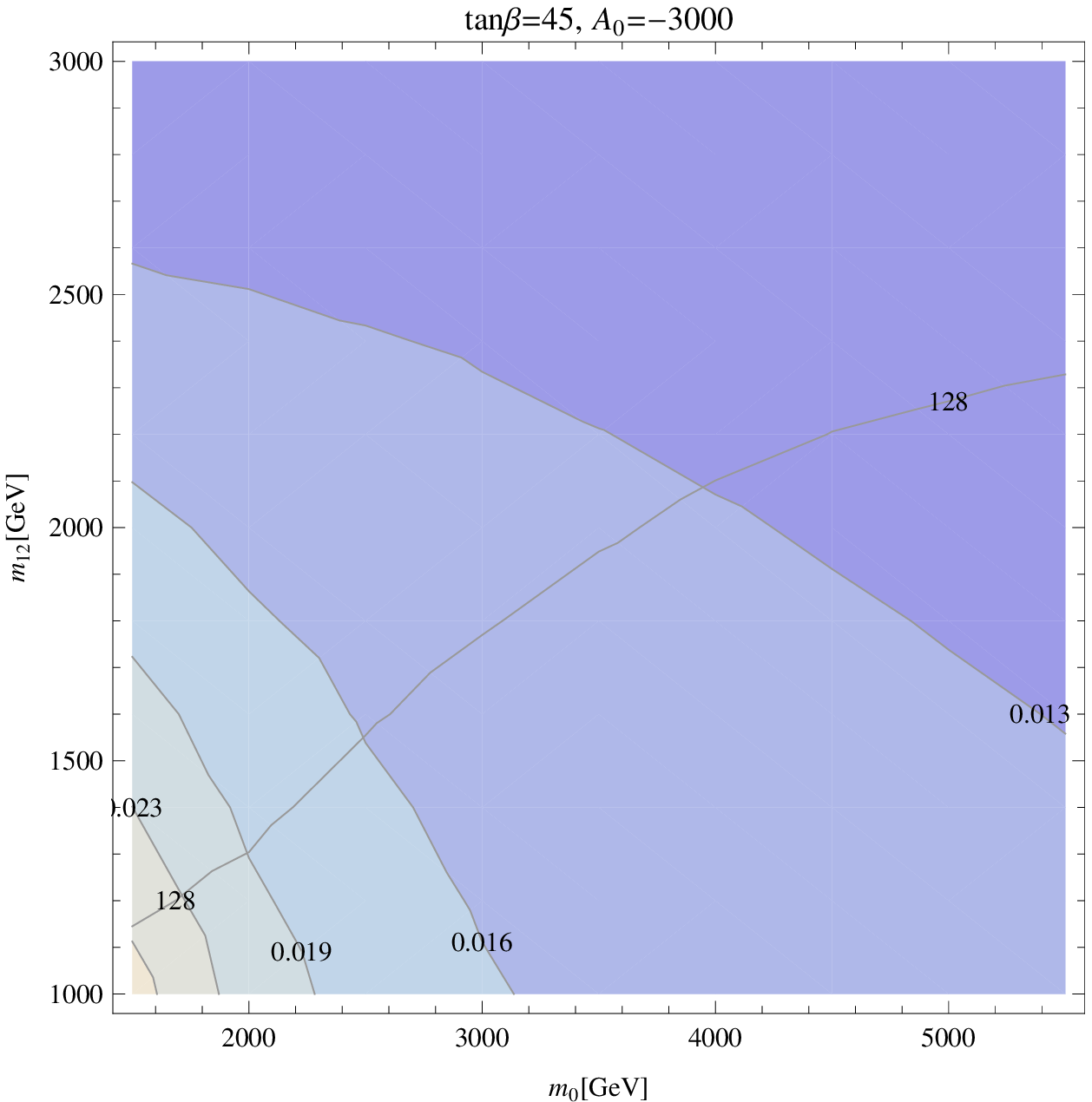   ,scale=0.51,angle=0,clip=}\\
\vspace{-0.2cm}
\end{center}
\caption{Contours of $\delta^{LLL}_{13}$ (Left) and $\delta^{LLL}_{23}$ (Right) in the
  $m_0$--$m_{1/2}$ plane in the \CMSSMI.  The line labeled as 128, correponds to a prediction $M_h=128$~GeV (see text), on the area above this line the prediction for $M_h$ is higher. } 
\label{fig:DelLLL23}
\end{figure}

Our numerical results in the \CMSSMI\ are shown in
\reffis{fig:DelLLL23} - \ref{fig:HTauMueSSI}. We have checked numerically that the dependence on
$\tb$ is not very prominent, but going from $A_0 = 0$ to $-3000 \gev$
has a strong impact on the $\deFABij$. For small $A_0$ the size of the
$\deFABij$ is increasing with larger $m_0$ and $m_{1/2}$, for 
$A_0 = -3000 \gev$ the largest values are found for small $m_0$ and
$m_{1/2}$. We present the
results in the $m_0$--$m_{1/2}$ plane for  
$\tb = 45$ and $A_0 = -3000 \gev$ only, capturing the ``largest'' case.
We start presenting the two most relevant $\deFABij$. 
Left plot in \reffis{fig:DelLLL23} show $\del{LLL}{13}$, 
right plot show $\del{LLL}{23}$. As expected,
$\del{LLL}{23}$ turns out to be largest of \order{0.01}, while the
$\del{LLL}{13}$ is one order of magnitude smaller. Contraints imposed by the Higgs mass are displayed on the plots, the areas above the line corresponding 
to $M_h=128$~GeV and below  $M_h=122$~GeV are excluded. Here we do not impose the satisfaction of the Cosmological bounds on  neutralino relic density, because  this is only achieved  on a few selected areas of the plots (an updated review can be found in Ref.~ \cite{Olive:2016efh} and references therein).

\subsection{\boldmath{${\rm BR}(l_i \rightarrow l_j \gamma)$}}
\begin{figure}[ht!]
\begin{center}
\psfig{file=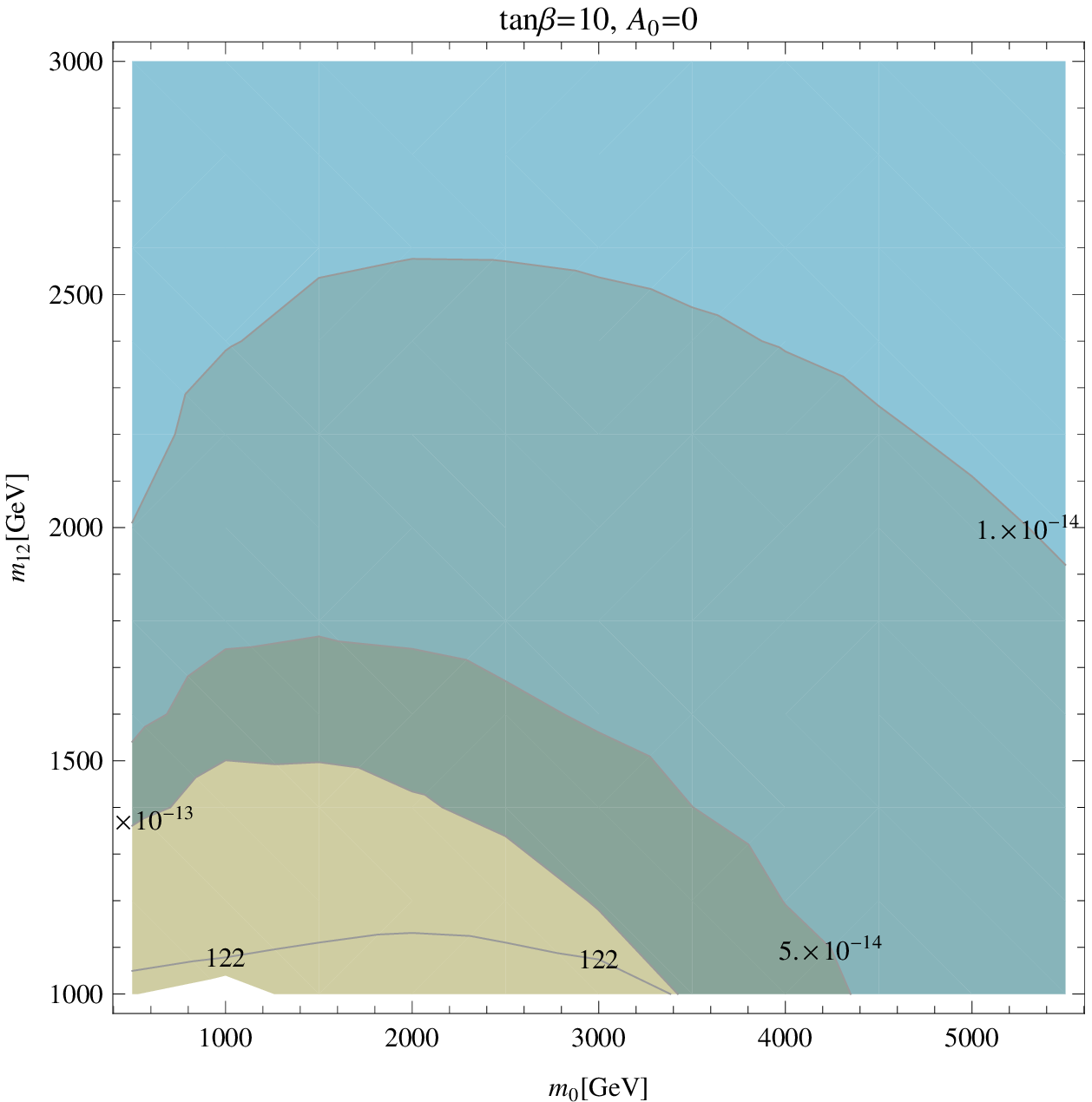  ,scale=0.51,angle=0,clip=}
\psfig{file=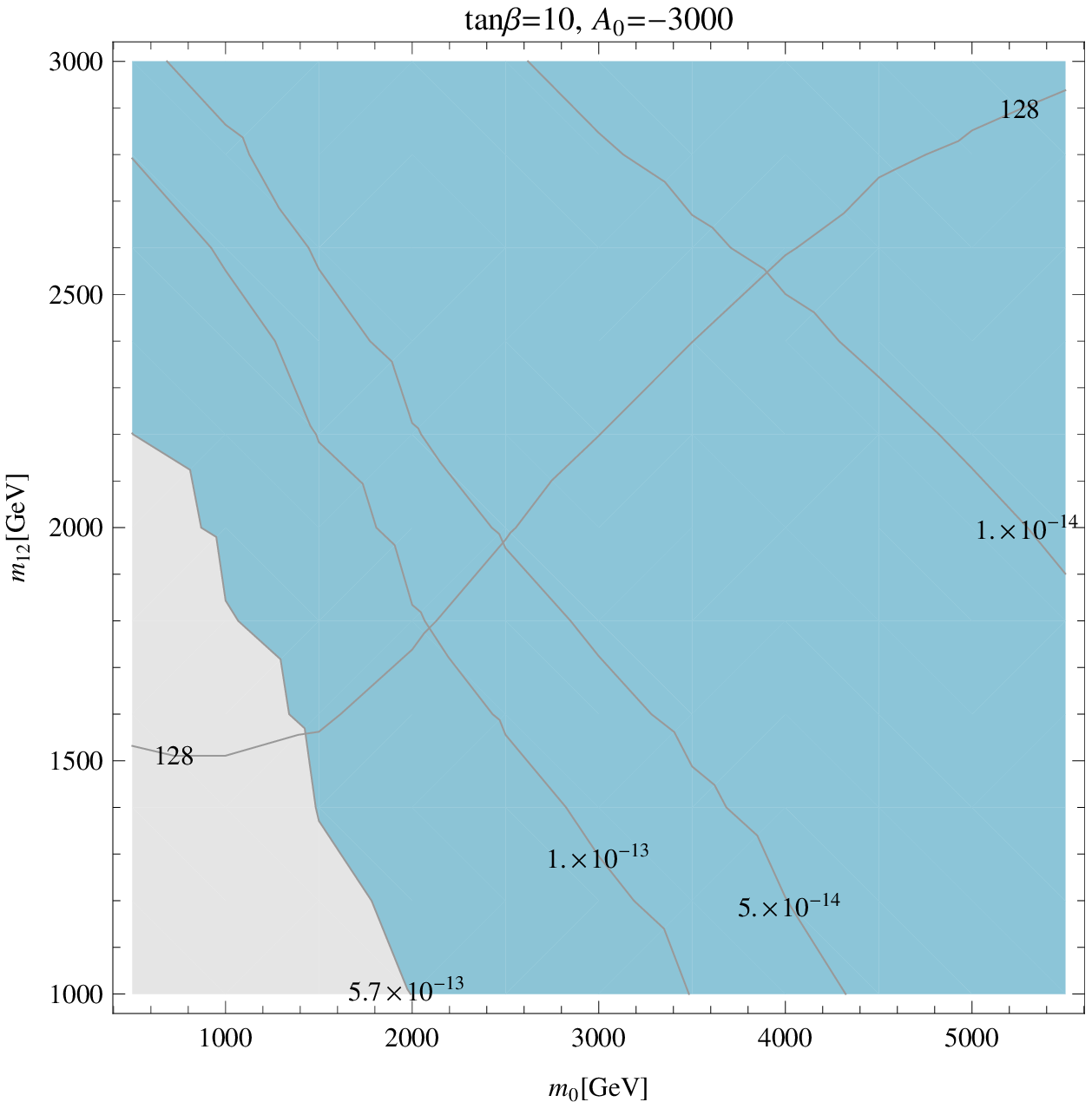  ,scale=0.51,angle=0,clip=}\\
\vspace{0.2cm}
\psfig{file=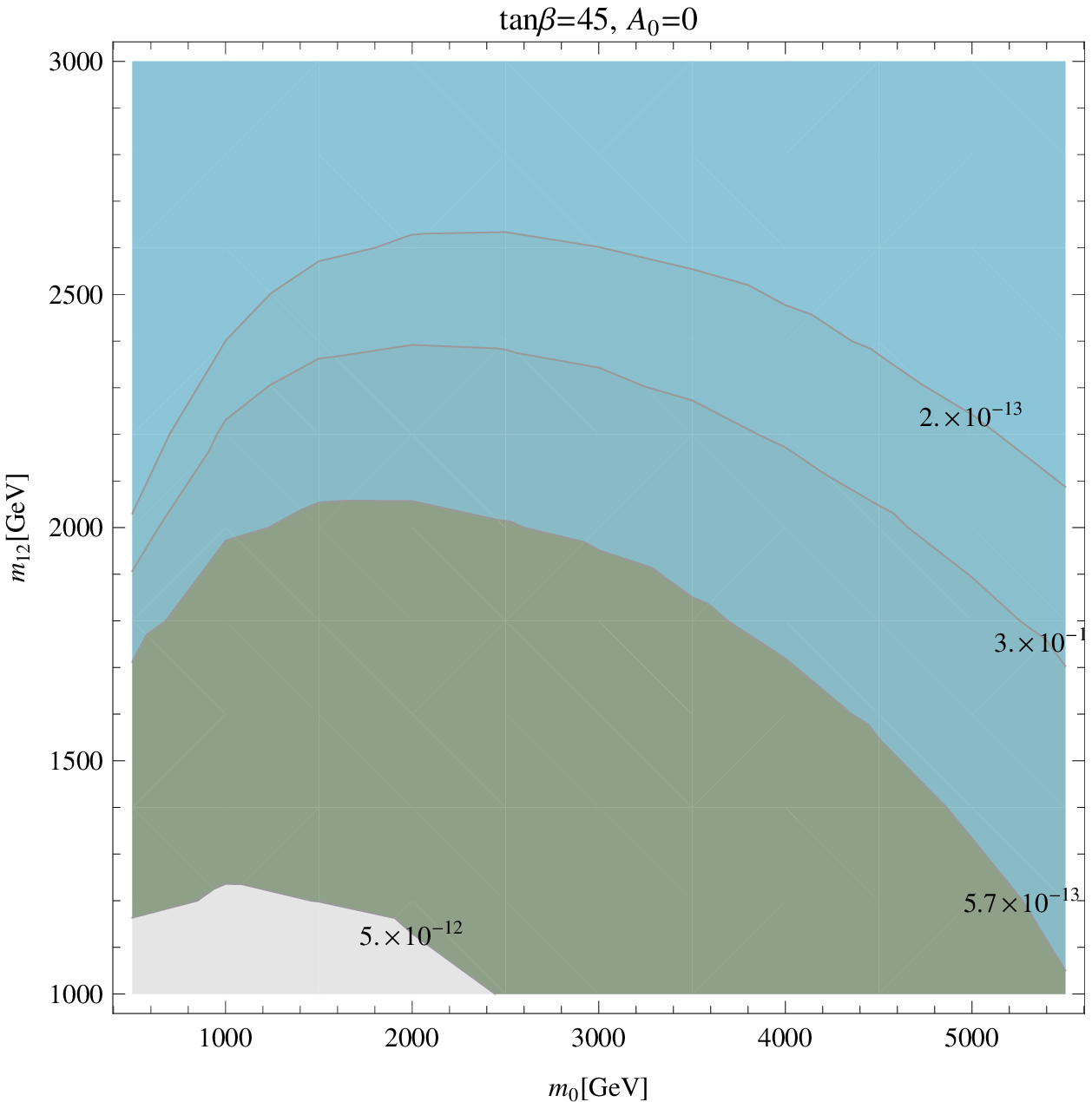 ,scale=0.51,angle=0,clip=}
\psfig{file=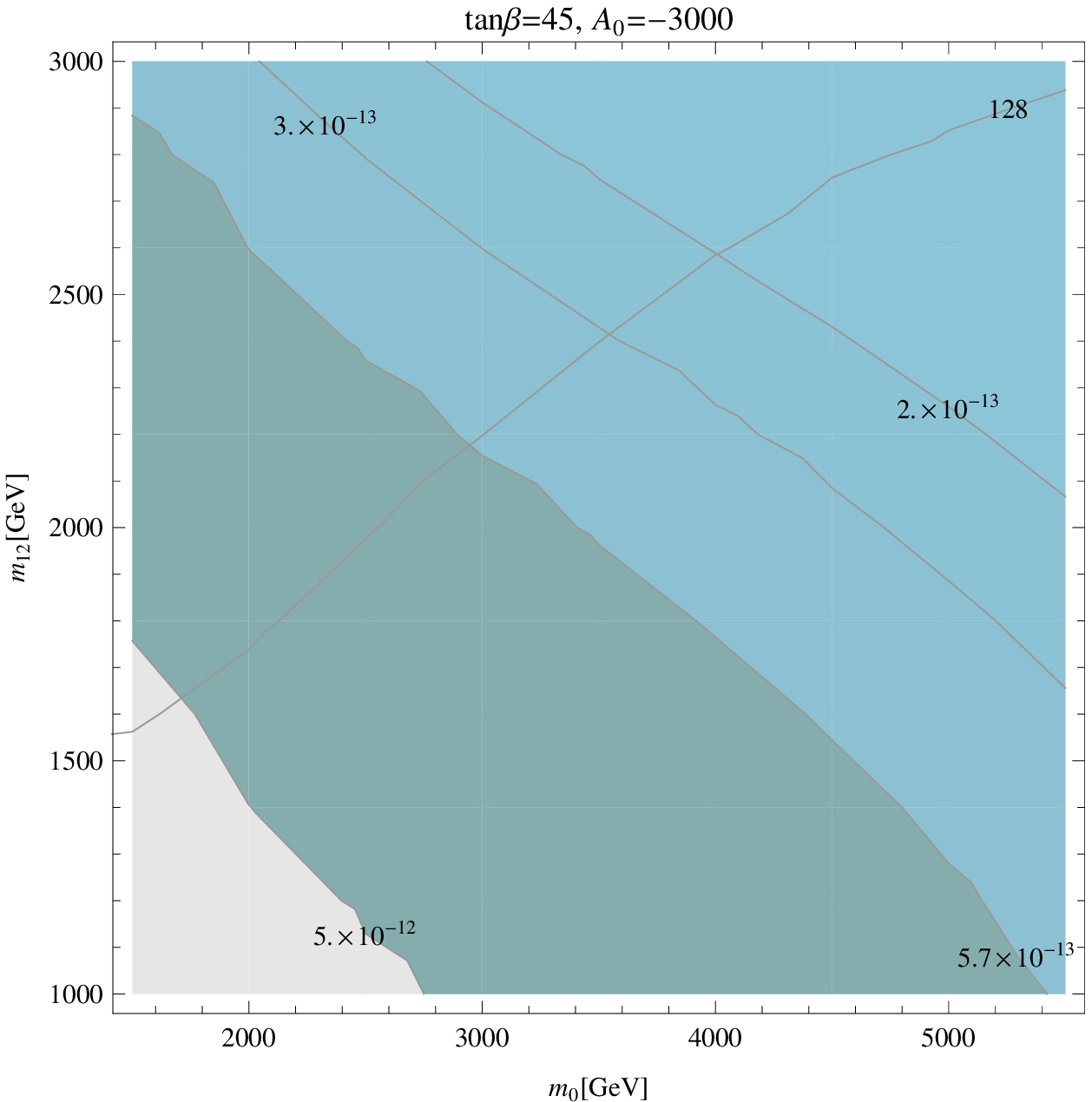   ,scale=0.51,angle=0,clip=}\\
\vspace{-0.2cm}
\end{center}
\caption[Contours of BR($\mu \to e \gamma$)  in the
  $m_0$--$m_{1/2}$ plane]{Contours of BR($\mu \to e \gamma$)  in the
  $m_0$--$m_{1/2}$ plane for different values of $\tb$ and $A_0$ in
  the \CMSSMI. The area below the  $5.7 \times 10^{-13}$ bound is excluded. The line labeled as 122 (128) on the plots of the left (right), correponds to a prediction $M_h=122 (128)$~GeV (see text), for the area below (above) this line the prediction for $M_h$ is lower (higher) than this value.}
\label{fig:BrmegSSI}
\end{figure}

\begin{figure}[ht!]
\begin{center}
\psfig{file=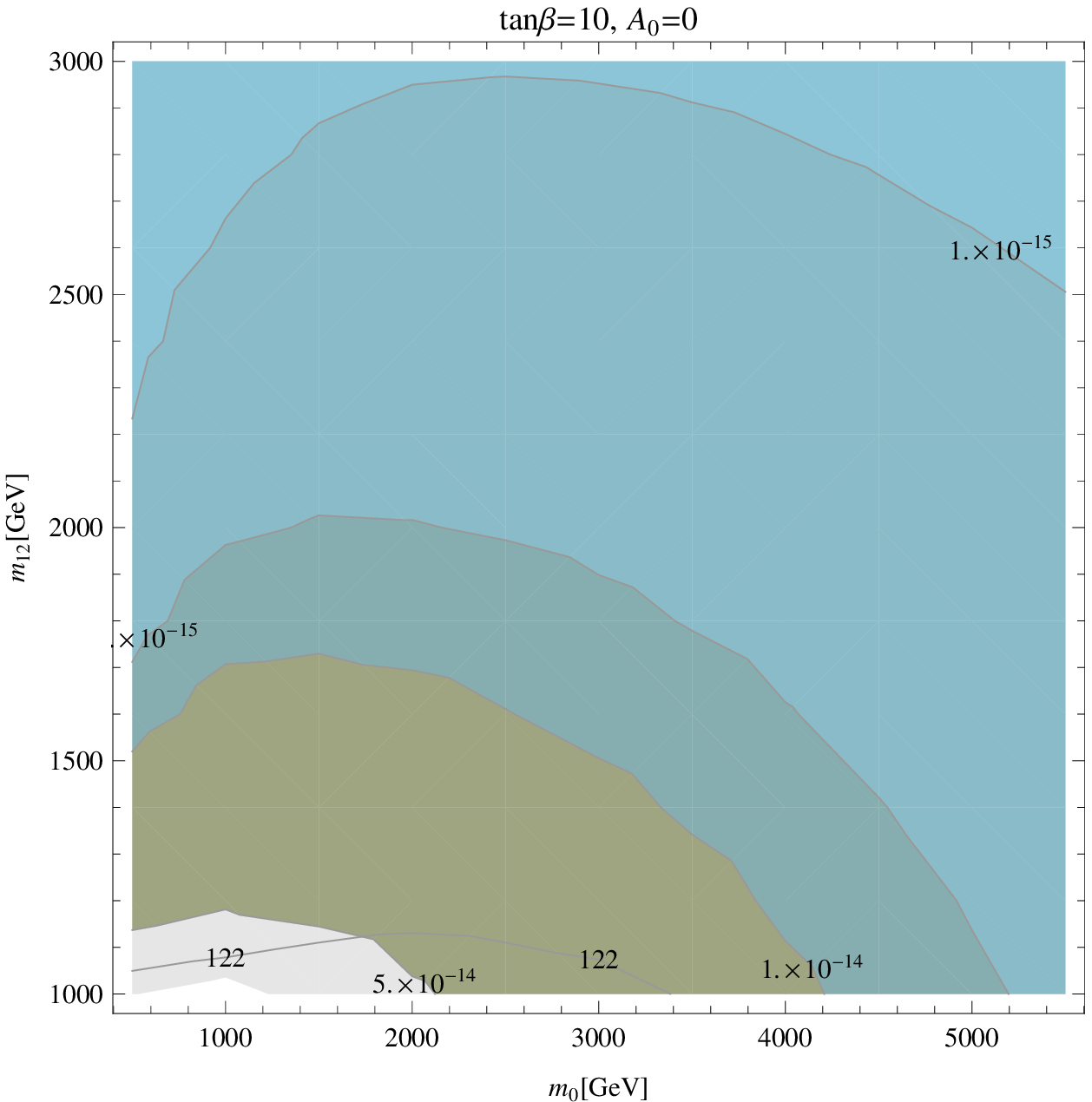  ,scale=0.51,angle=0,clip=}
\psfig{file=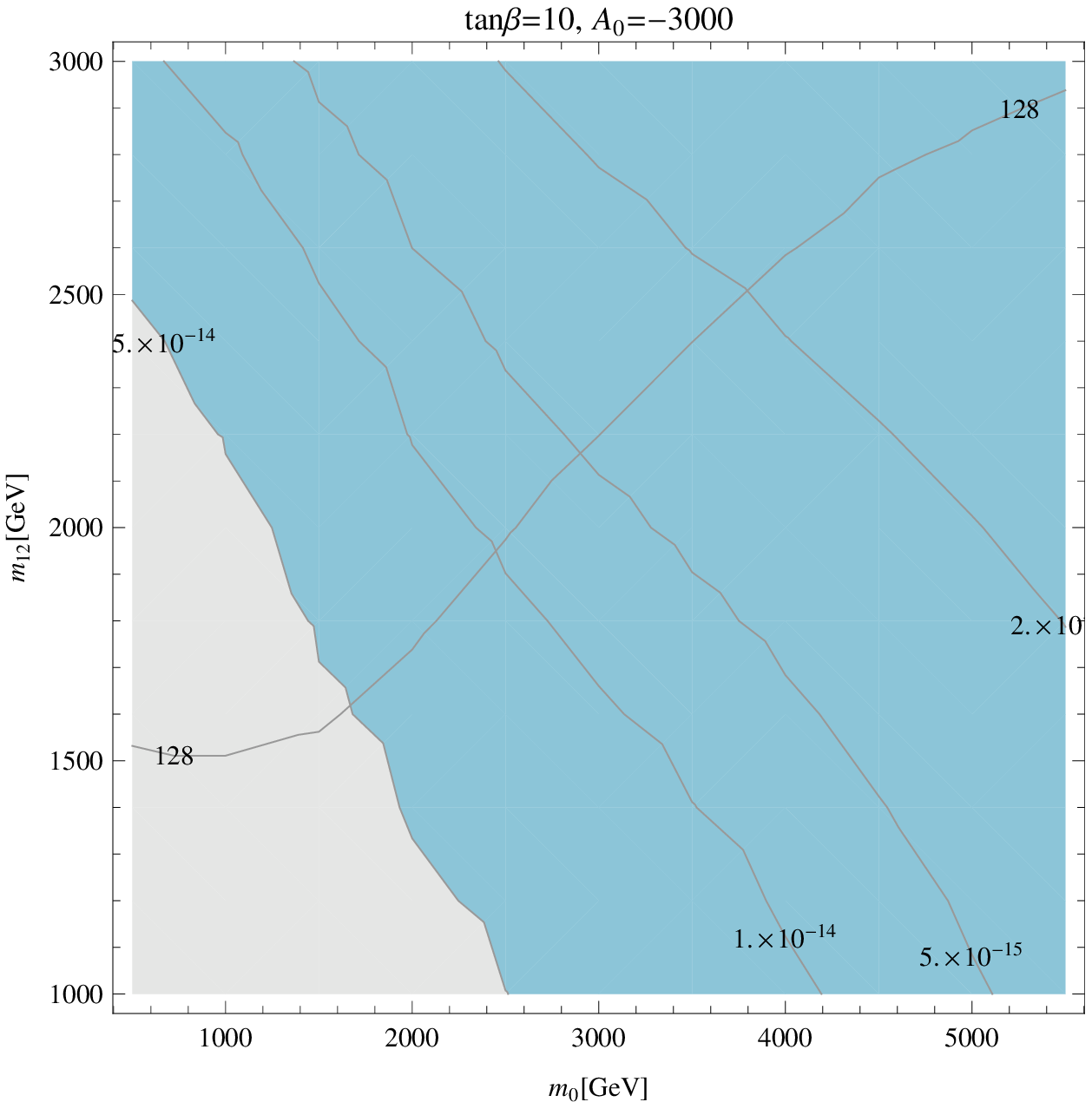  ,scale=0.51,angle=0,clip=}\\
\vspace{0.2cm}
\psfig{file=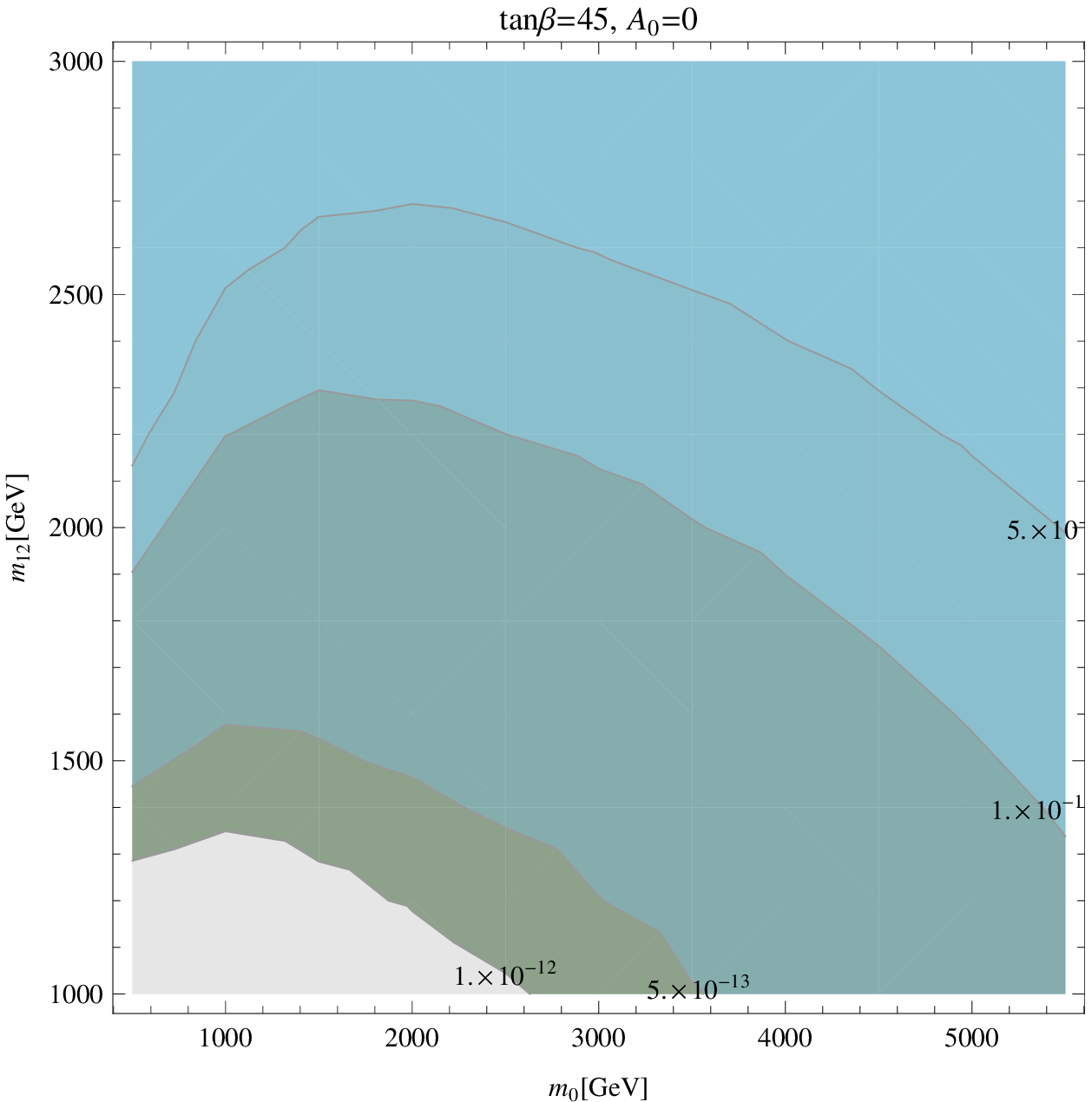 ,scale=0.51,angle=0,clip=}
\psfig{file=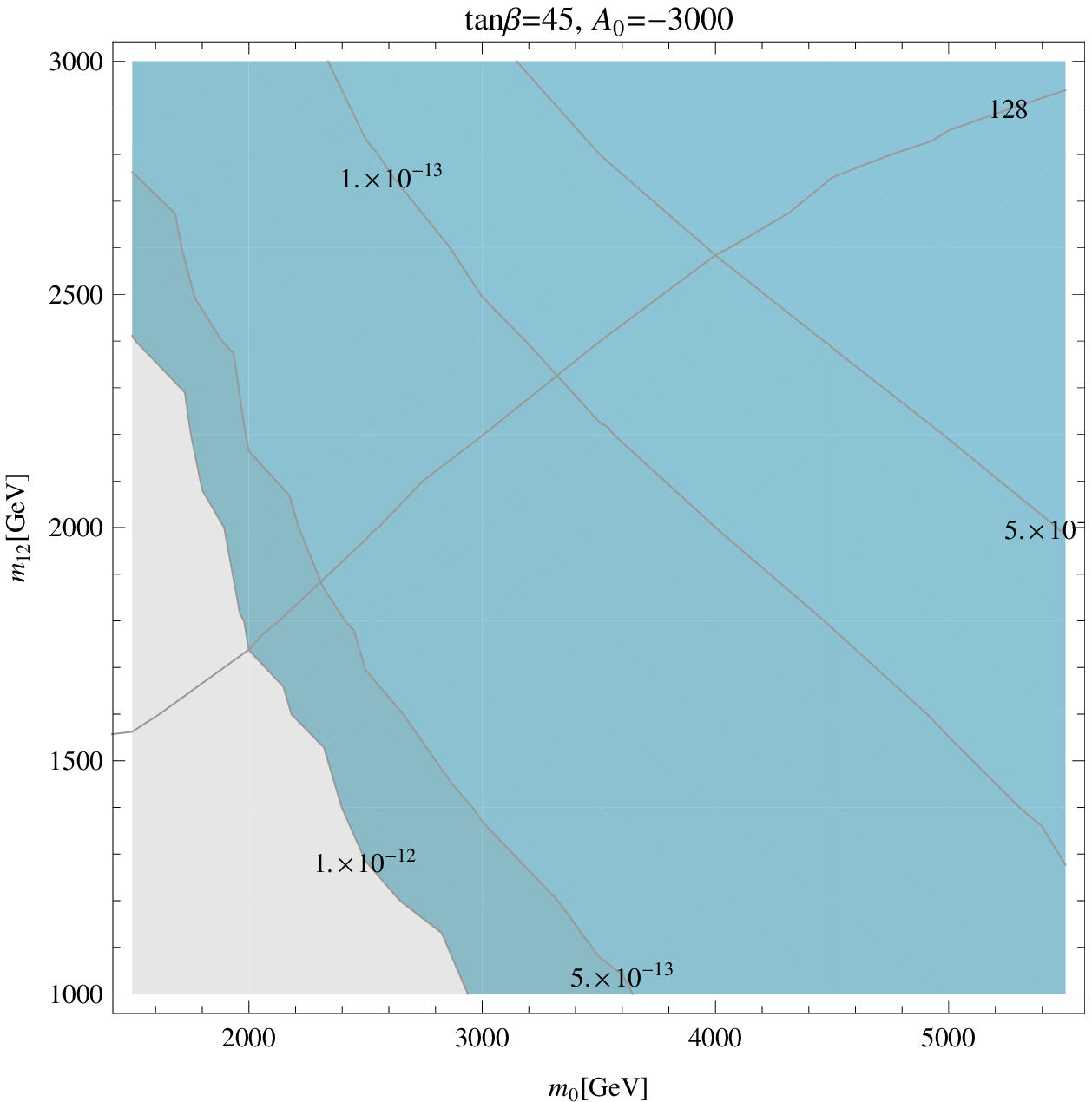   ,scale=0.51,angle=0,clip=}\\
\vspace{-0.2cm}
\end{center}
\caption[Contours of BR($\tau \to e \gamma$)  in the
  $m_0$--$m_{1/2}$ plane]{Contours of BR($\tau \to e \gamma$)  in the
  $m_0$--$m_{1/2}$ plane for different values of $\tb$ and $A_0$ in
  the \CMSSMI. Lines labeled as 122 and 128 are as descibred in Figs.\ref{fig:DelLLL23} and \ref{fig:BrmegSSI}}   
\label{fig:BrtegSSI}
\end{figure}

\begin{figure}[ht!]
\begin{center}
\psfig{file=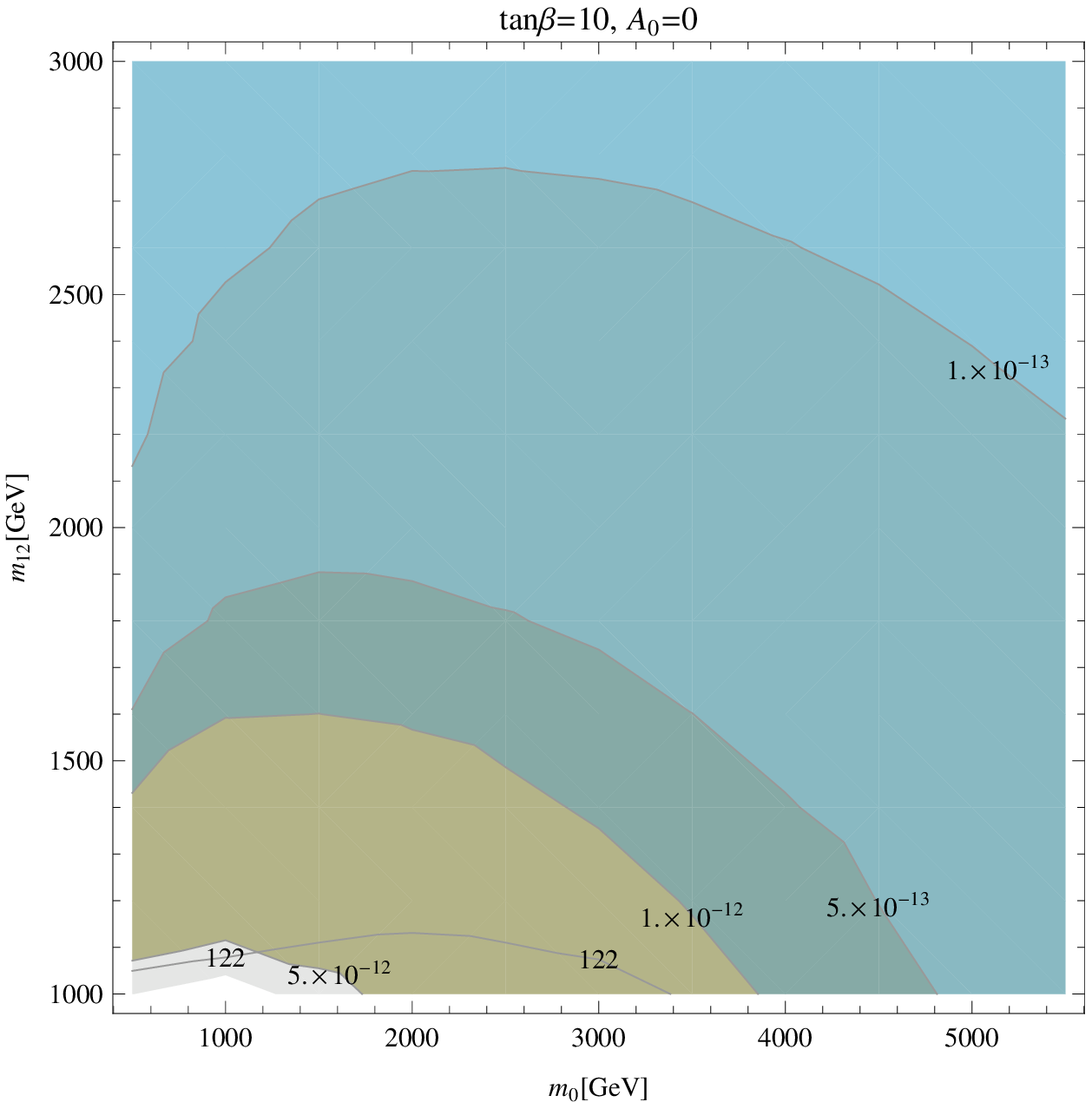  ,scale=0.51,angle=0,clip=}
\psfig{file=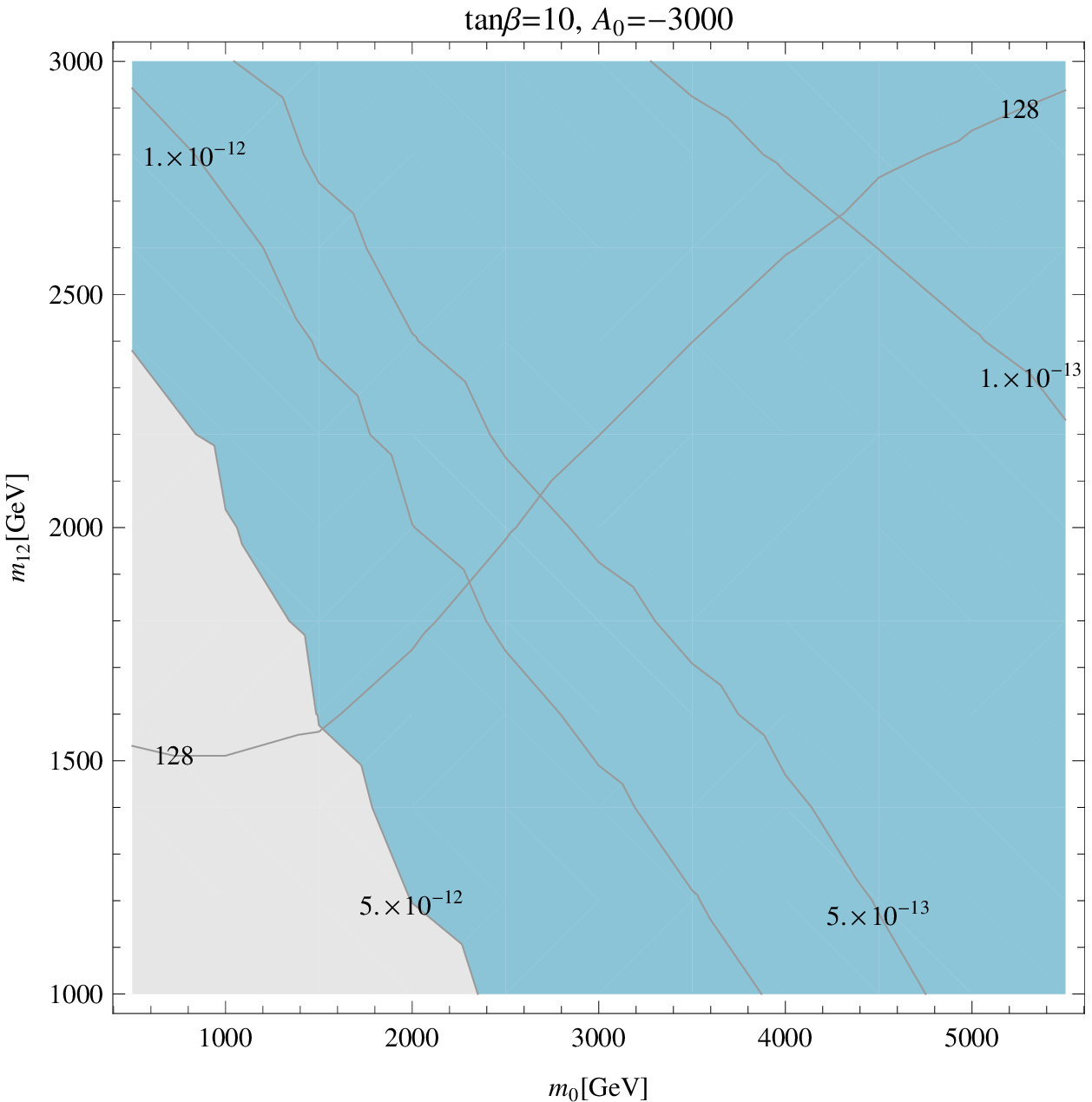  ,scale=0.51,angle=0,clip=}\\
\vspace{0.2cm}
\psfig{file=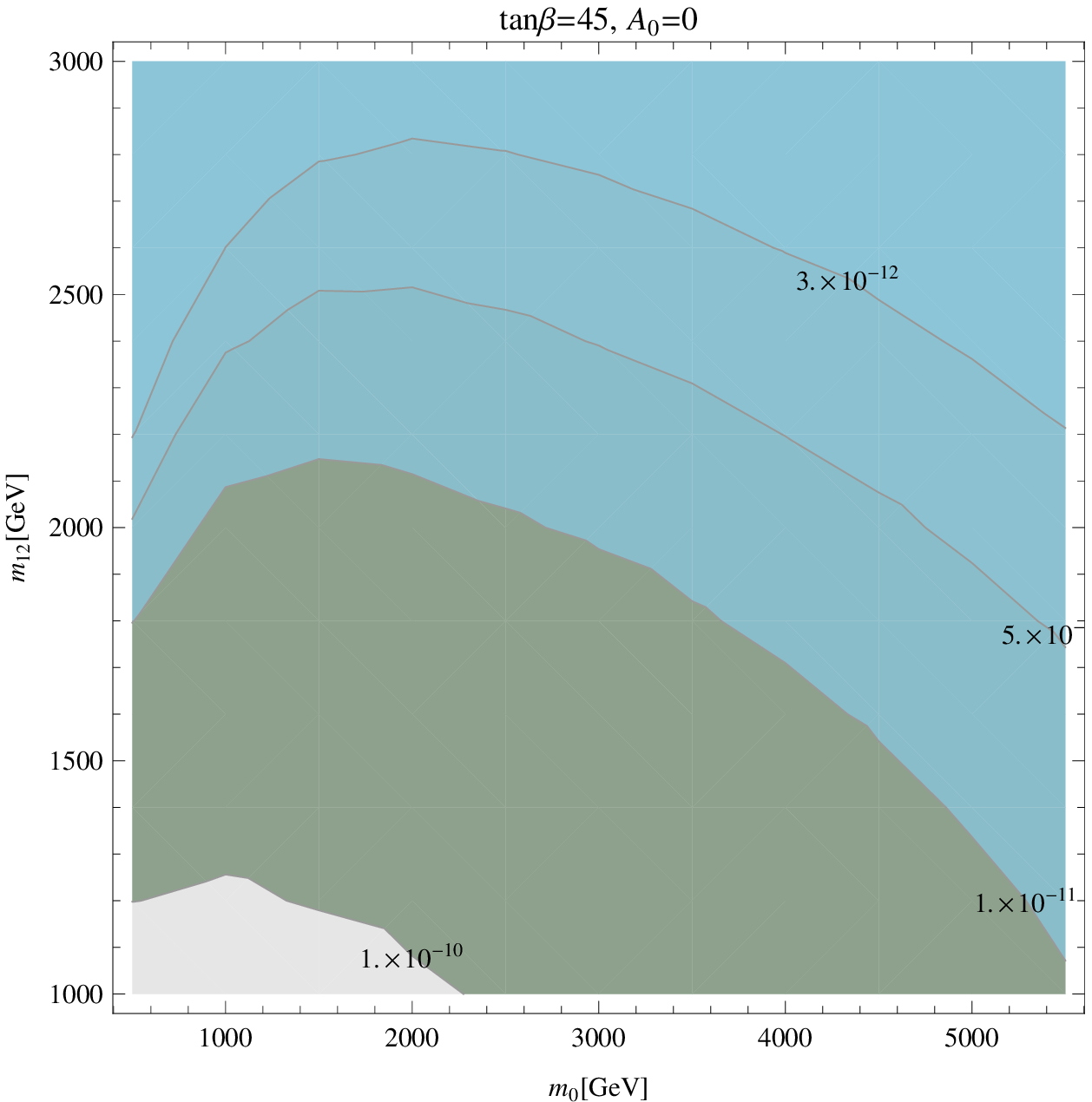 ,scale=0.51,angle=0,clip=}
\psfig{file=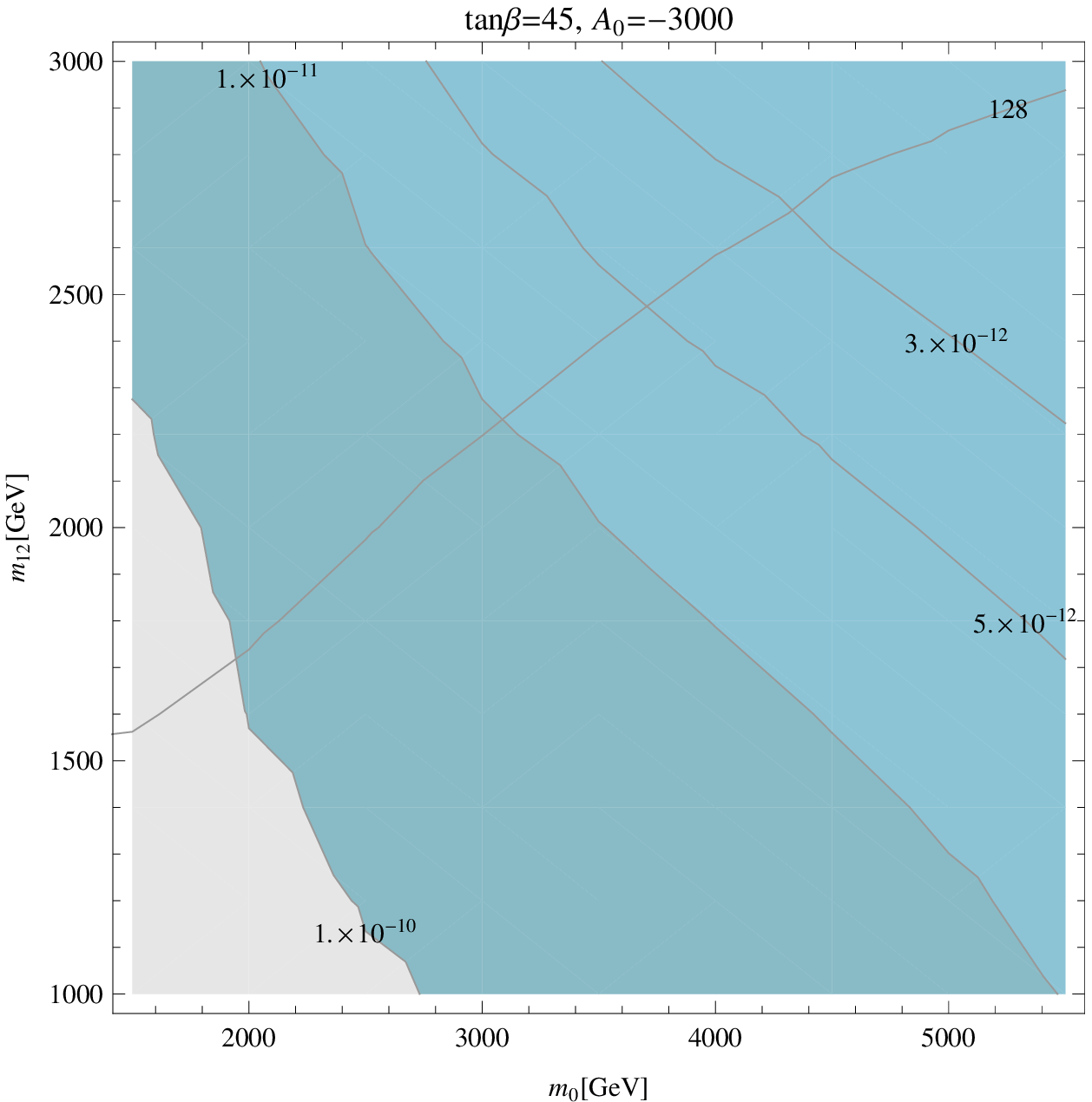   ,scale=0.51,angle=0,clip=}\\
\vspace{-0.2cm}
\end{center}
\caption[Contours of BR($\tau \to \mu \gamma$)  in the
  $m_0$--$m_{1/2}$ plane]{Contours of BR($\tau \to \mu \gamma$)  in the
  $m_0$--$m_{1/2}$ plane for different values of $\tb$ and $A_0$ in
  the \CMSSMI. Lines labeled as 122 and 128 are as descibred in Figs.\ref{fig:DelLLL23} and \ref{fig:BrmegSSI}}   
\label{fig:BrtmgSSI}
\end{figure}

The experimental limit BR($\mu \to e \gamma)< 5.7 \times 10^{-13}$ put severe constraints on slepton $\deFABij$'s as discussed before.
In \reffi{fig:BrmegSSI}, we show the predictions for BR($\mu \to e \gamma$)
in $m_0$--$m_{1/2}$ plane for different values of $A_0$ and $\tb$ in \CMSSMI.
The selected values of $Y_\nu$ result in a large prediction for, e.g.,
BR($\mu \to e \gamma$) that can eliminate some of the $m_0$--$m_{1/2}$
parameter plane, in particular combinations of low values of $m_0$ and
$m_{1/2}$. For $\tb=10$ and $A_0=0$, BR($\mu \to e \gamma$) (upper left plot of \reffi{fig:BrmegSSI}) do not exclude any region in $m_0$--$m_{1/2}$ plane, whereas with $\tb=10$ and $A_0=-3000$ lower left region below $m_0, m_{1/2}=2000 $ is excluded (see upper right plot of \reffi{fig:BrmegSSI}).  For combinations like $\tb=45,$ $A_0=0$ and $\tb=45,$ $A_0=-3000$ even larger parts of the plane are excluded by BR($\mu \to e \gamma$).
In \reffi{fig:BrtegSSI} and \reffi{fig:BrtmgSSI}, we show the predictions for BR($\tau \to e \gamma$) and BR($\tau \to \mu \gamma$) respectively.  It can be seen that these processes do not reach their respective experimental bounds ${\rm BR}(\tau \rightarrow e \gamma)< 3.3 \times 10^{-8}$, ${\rm BR}(\tau \rightarrow \mu \gamma)< 4.4 \times 10^{-8}$. Consequently they do not exclude any parameter space.

\subsection{\boldmath${\rm BR}(h \rightarrow l_i^{\pm} l_j^{\mp}$)}

\begin{figure}[ht!]
\begin{center}
\psfig{file=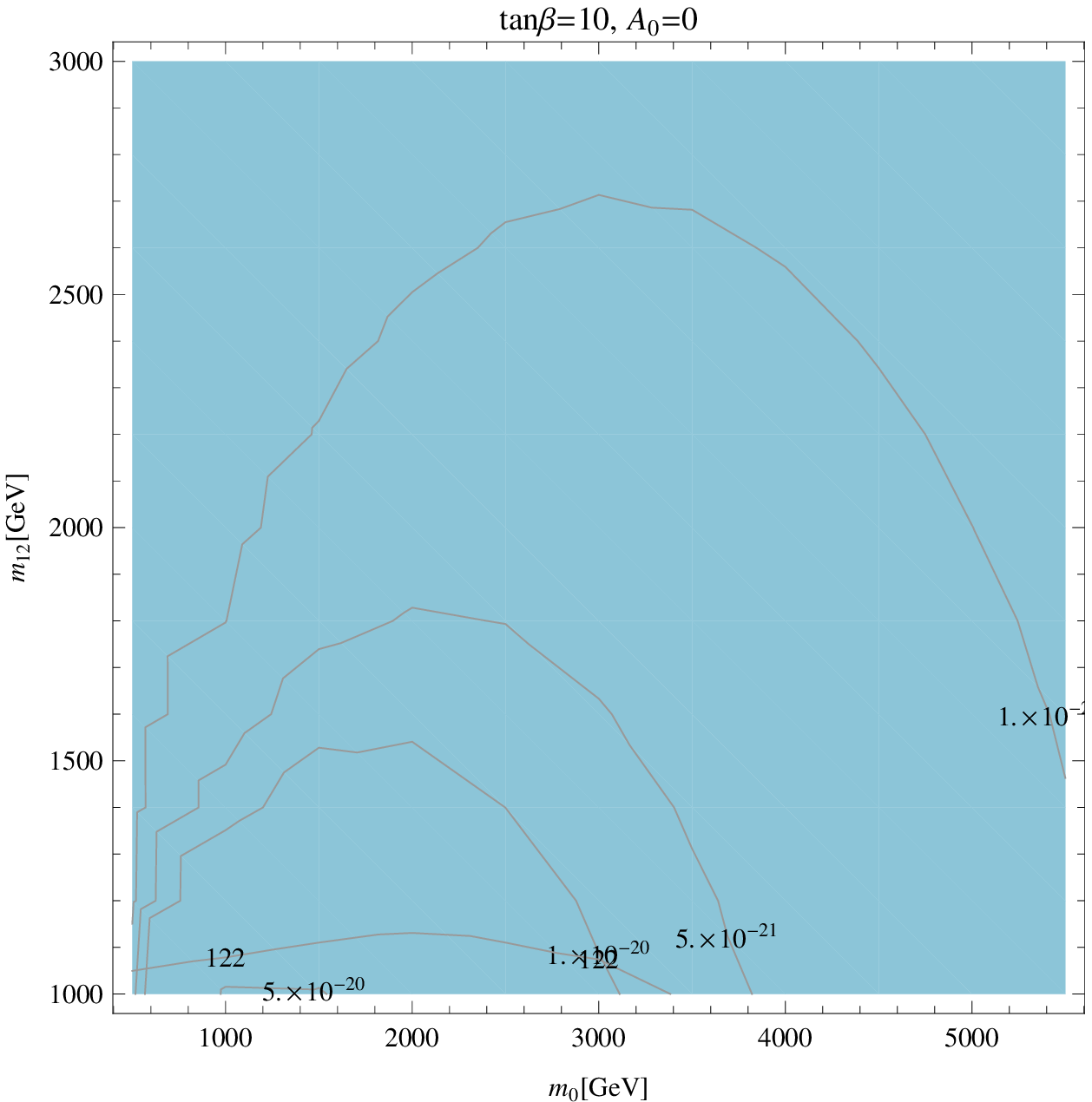  ,scale=0.51,angle=0,clip=}
\psfig{file=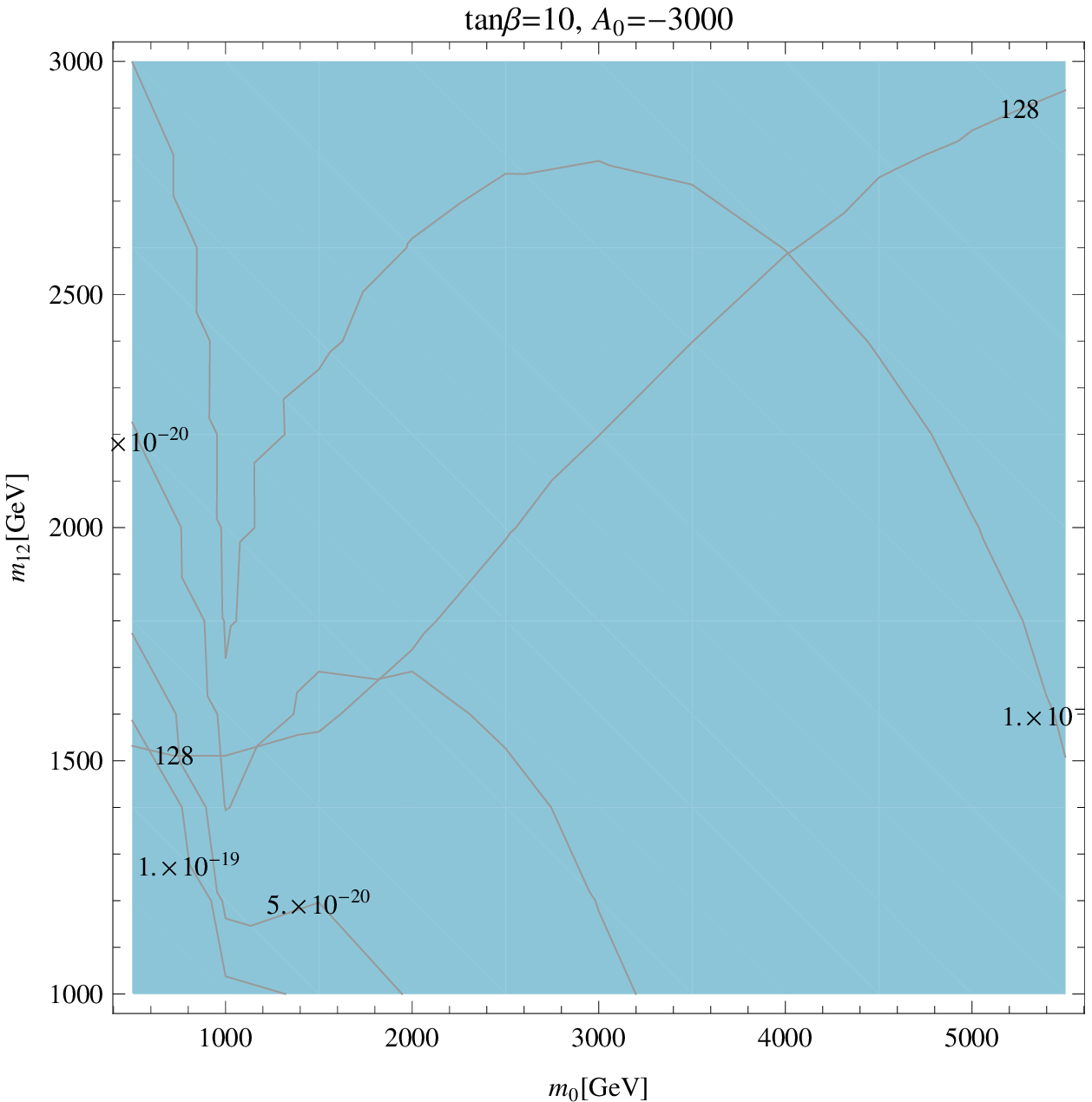  ,scale=0.51,angle=0,clip=}\\
\vspace{0.2cm}
\psfig{file=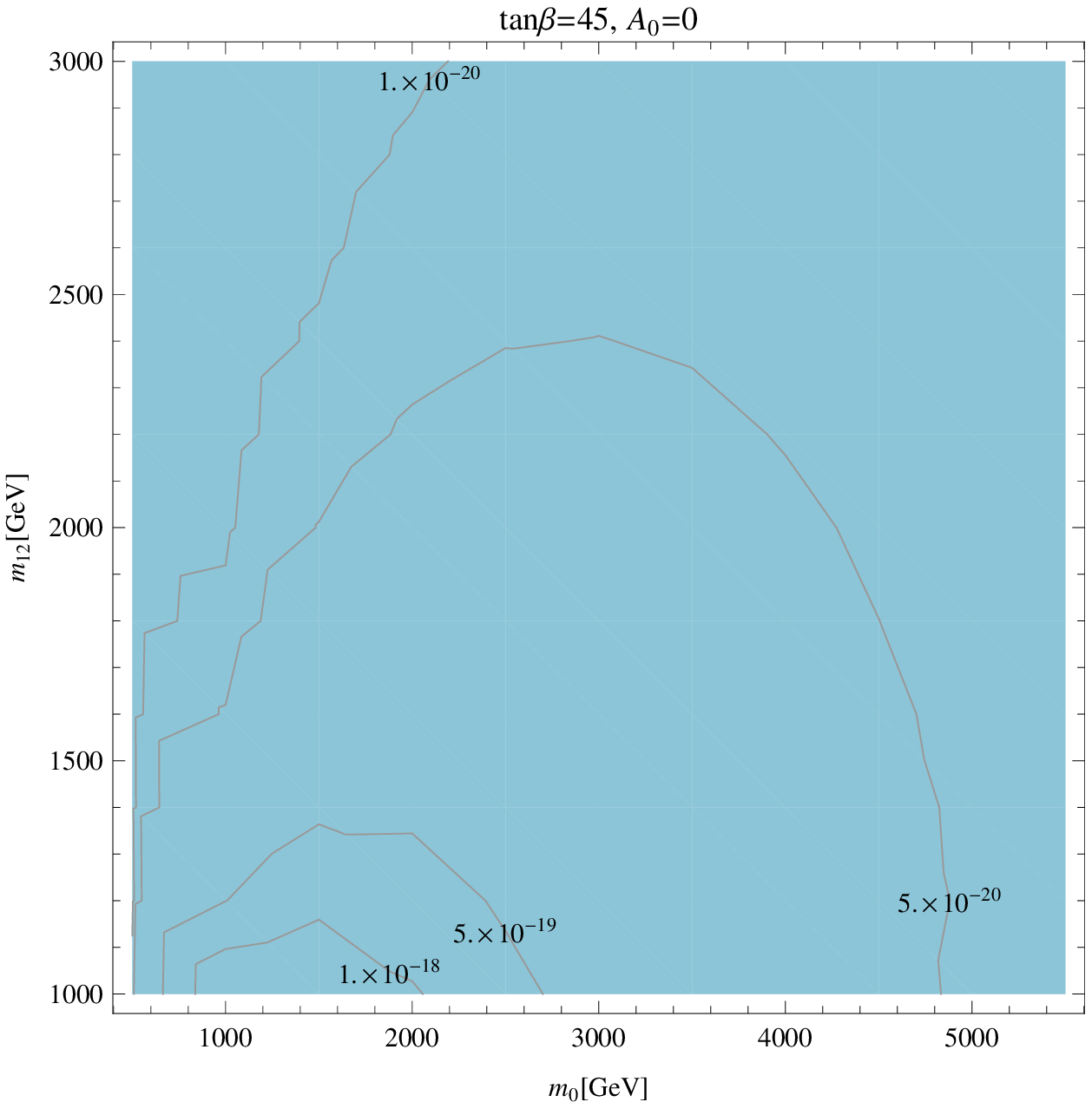 ,scale=0.51,angle=0,clip=}
\psfig{file=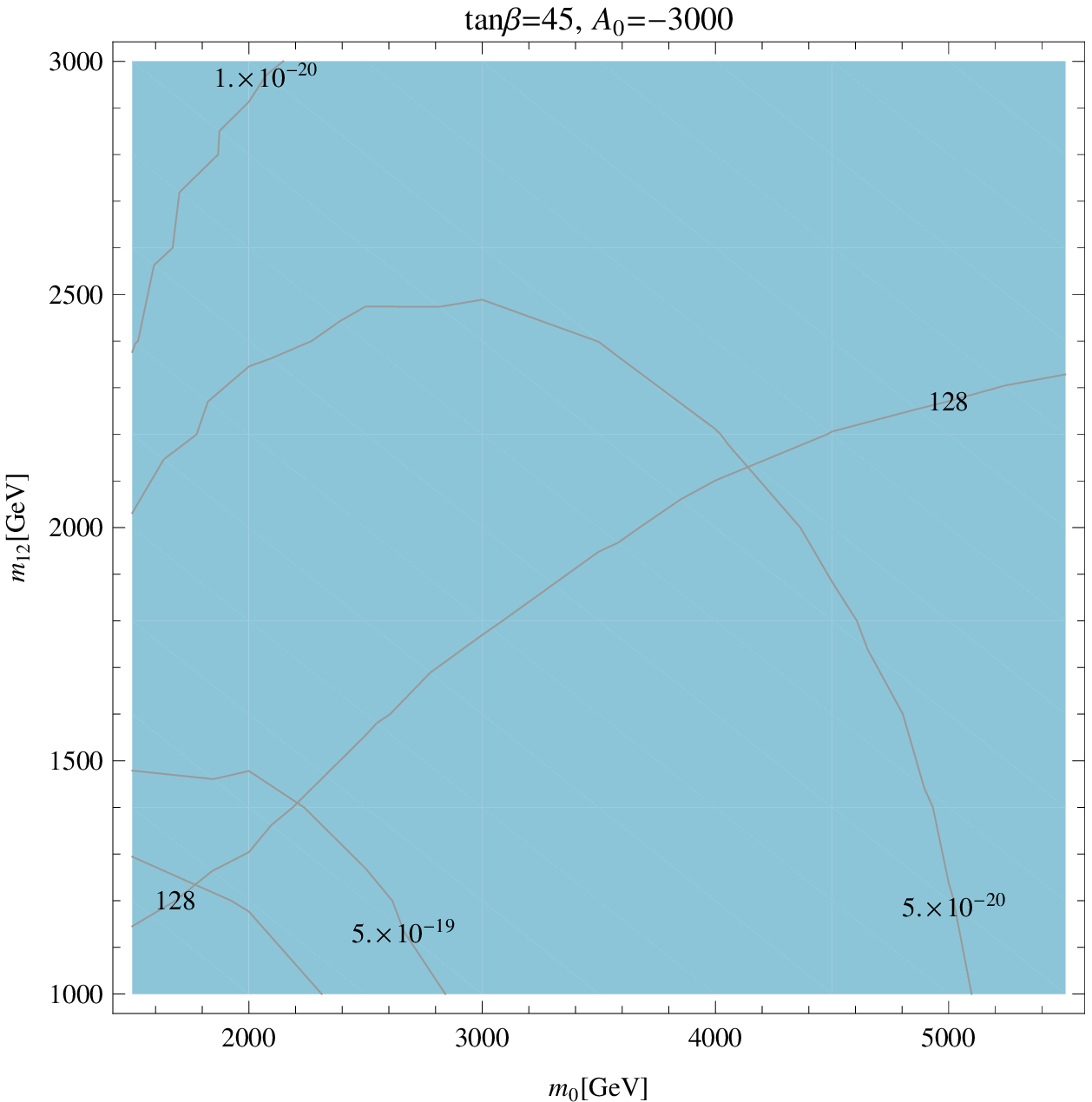   ,scale=0.51,angle=0,clip=}\\
\vspace{-0.2cm}
\end{center}
\caption[Contours of BR($h \to e \mu$)  in the
  $m_0$--$m_{1/2}$ plane]{Contours of BR($h \to e \mu$)  in the
  $m_0$--$m_{1/2}$ plane for different values of $\tb$ and $A_0$ in
  the \CMSSMI. Lines labeled as 122 and 128 are as descibred in Figs.\ref{fig:DelLLL23} and \ref{fig:BrmegSSI}}   
\label{fig:HMueESSI}
\end{figure}

\begin{figure}[ht!]
\begin{center}
\psfig{file=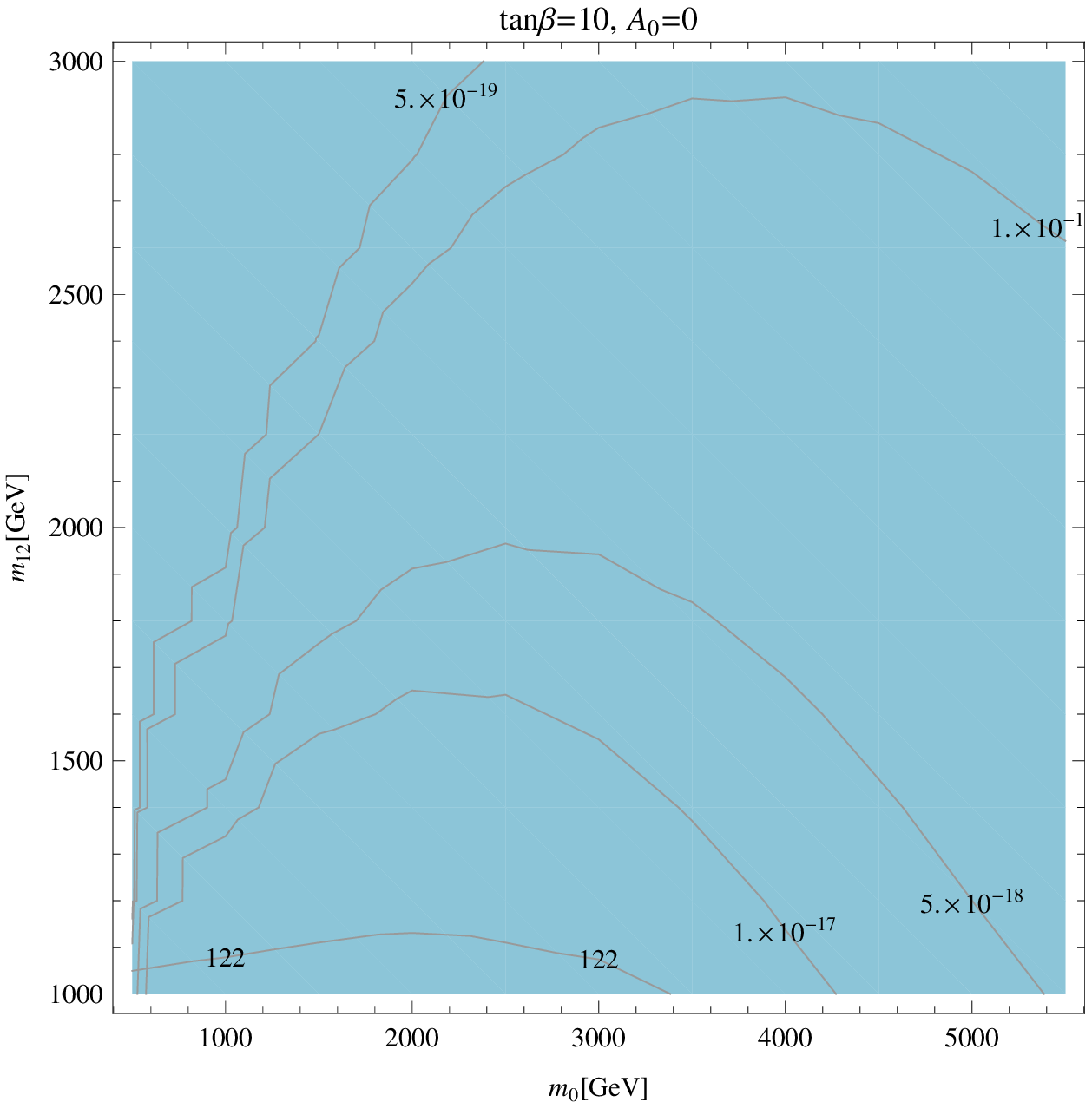  ,scale=0.51,angle=0,clip=}
\psfig{file=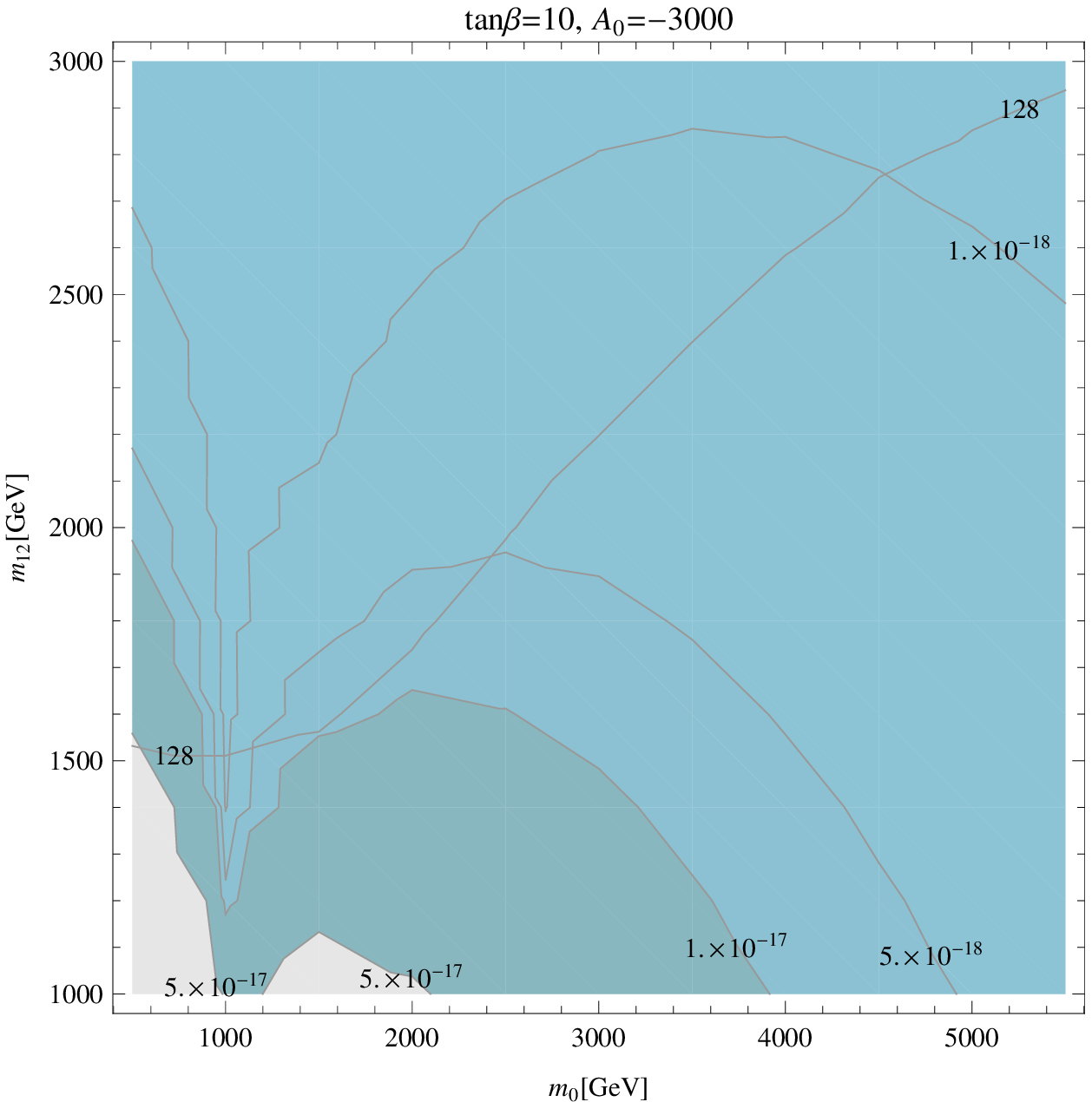  ,scale=0.51,angle=0,clip=}\\
\vspace{0.2cm}
\psfig{file=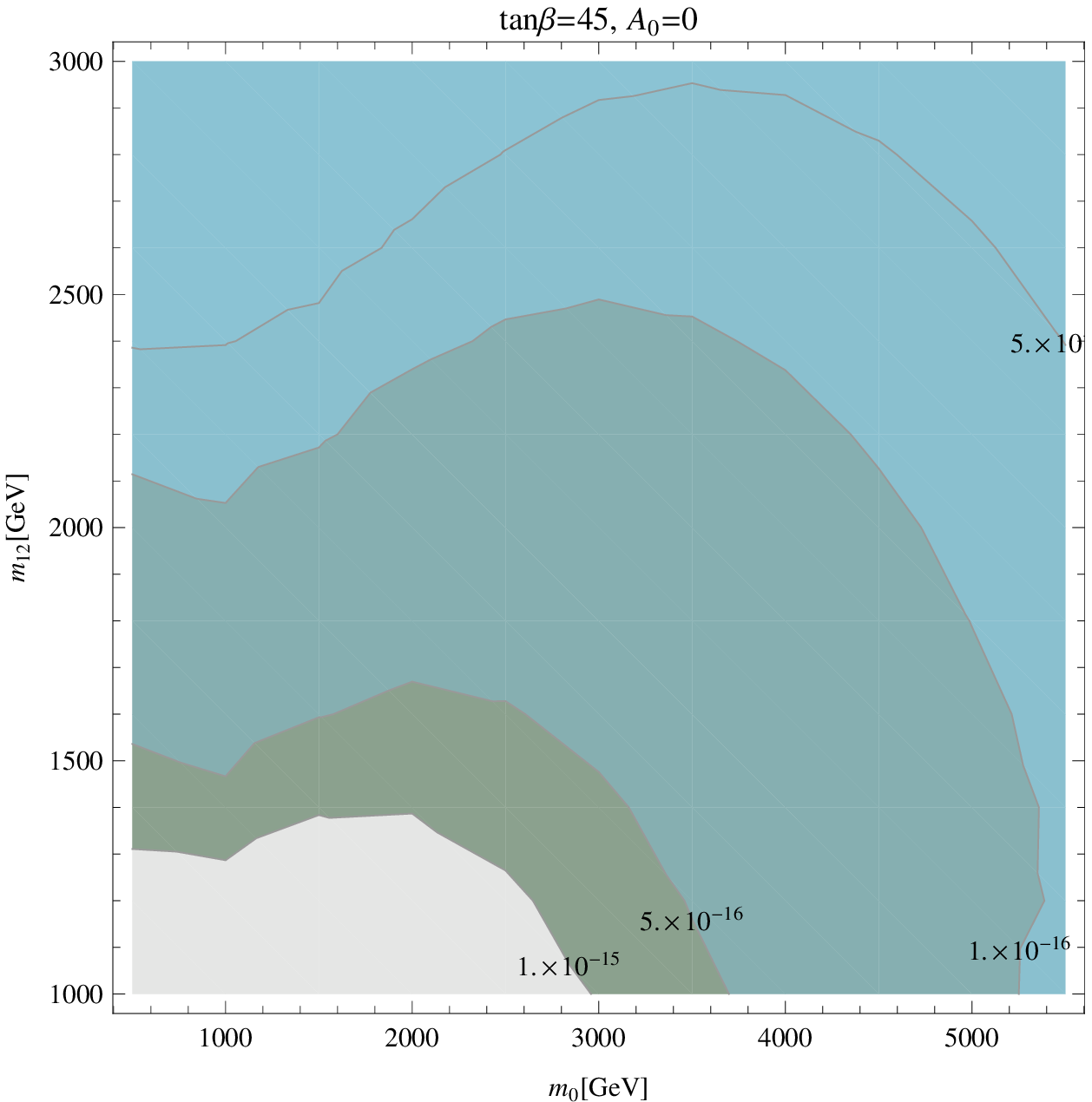 ,scale=0.51,angle=0,clip=}
\psfig{file=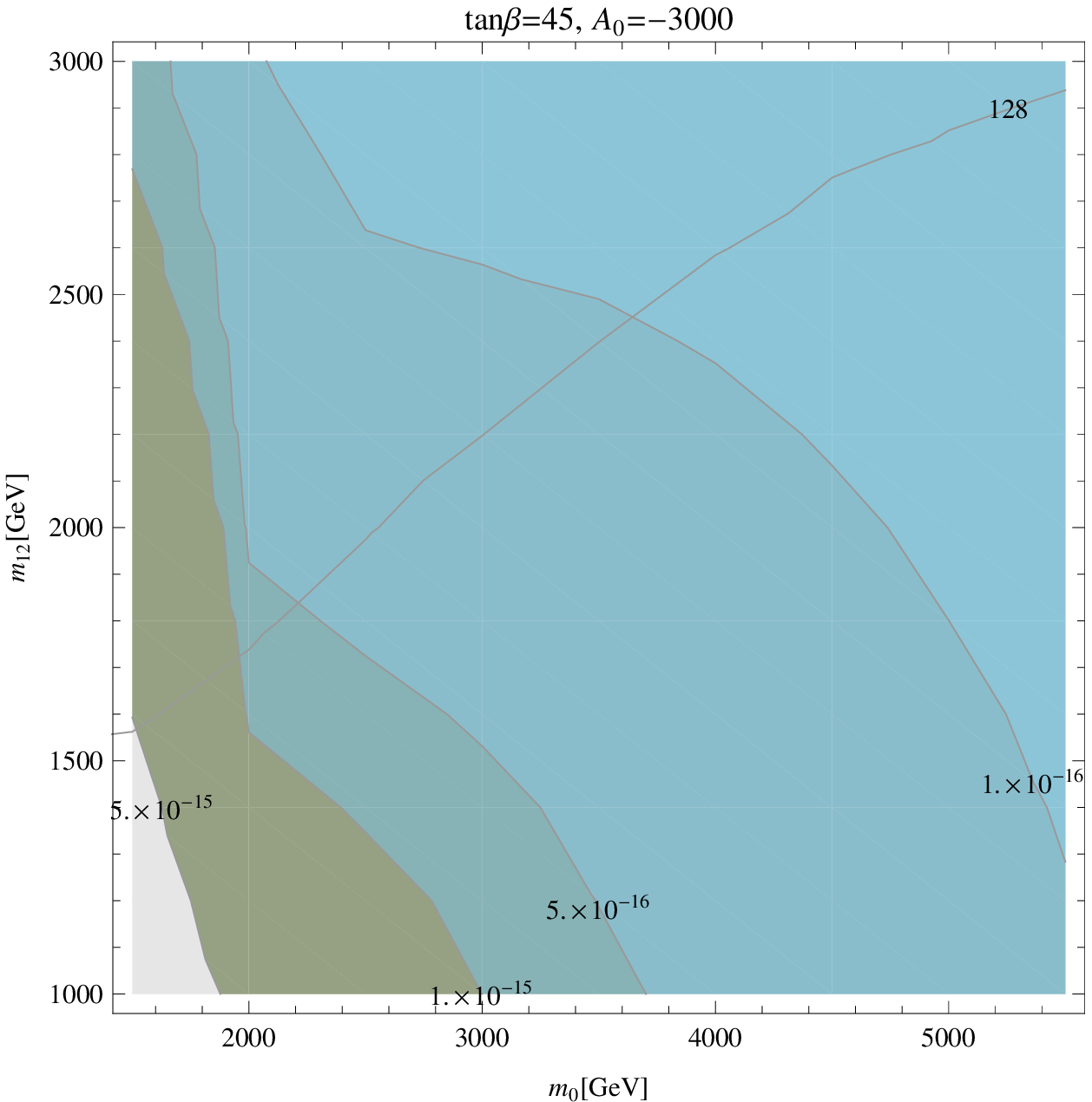   ,scale=0.51,angle=0,clip=}\\
\vspace{-0.2cm}
\end{center}
\caption[Contours of BR($h \to e \tau$)  in the
  $m_0$--$m_{1/2}$ plane]{Contours of BR($h \to e \tau$)  in the
  $m_0$--$m_{1/2}$ plane for different values of $\tb$ and $A_0$ in
  the \CMSSMI. Lines labeled as 122 and 128 are as descibred in Figs.\ref{fig:DelLLL23} and \ref{fig:BrmegSSI}}   
\label{fig:HTauESSI}
\end{figure}

\begin{figure}[ht!]
\begin{center}
\psfig{file=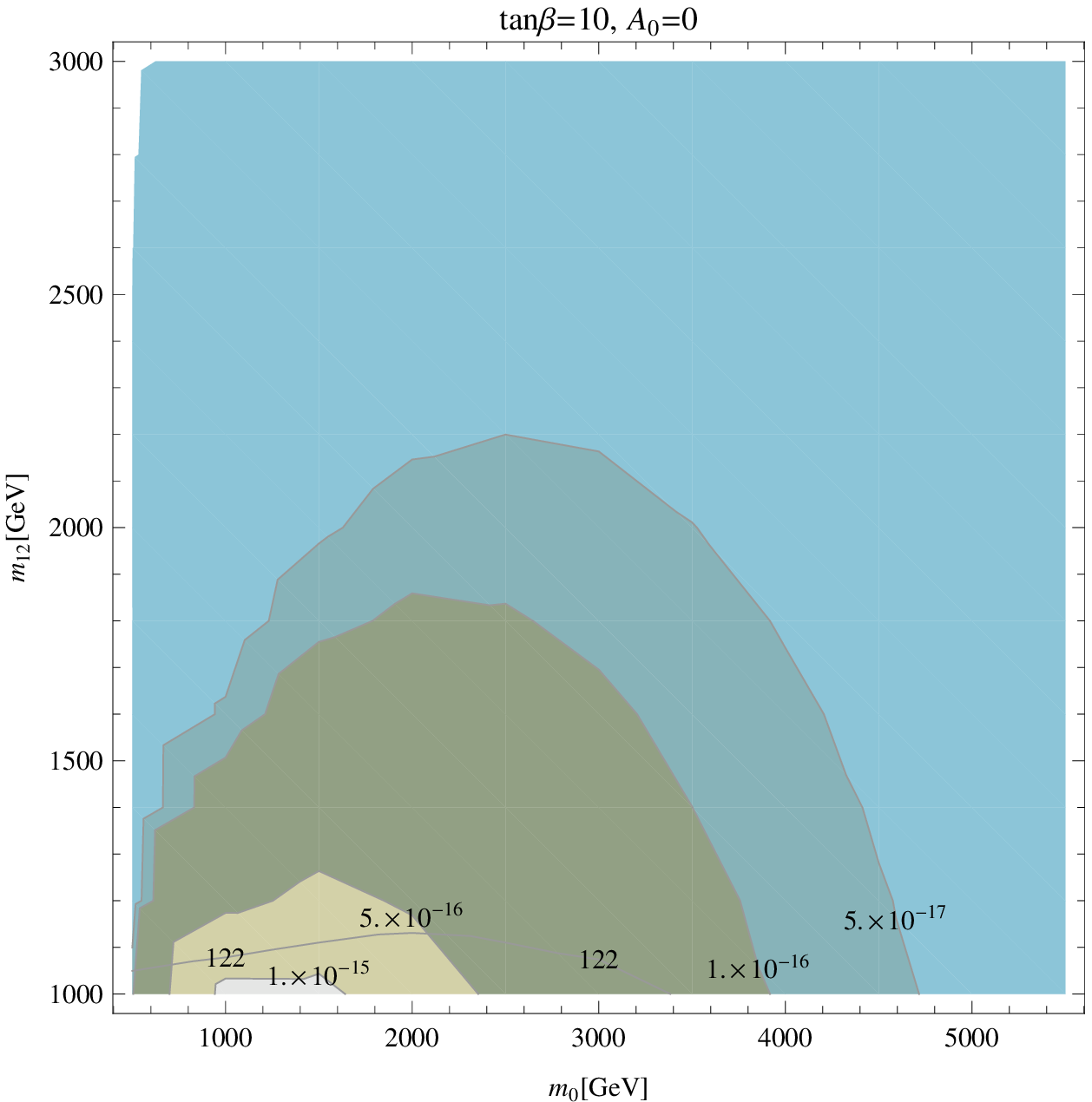  ,scale=0.51,angle=0,clip=}
\psfig{file=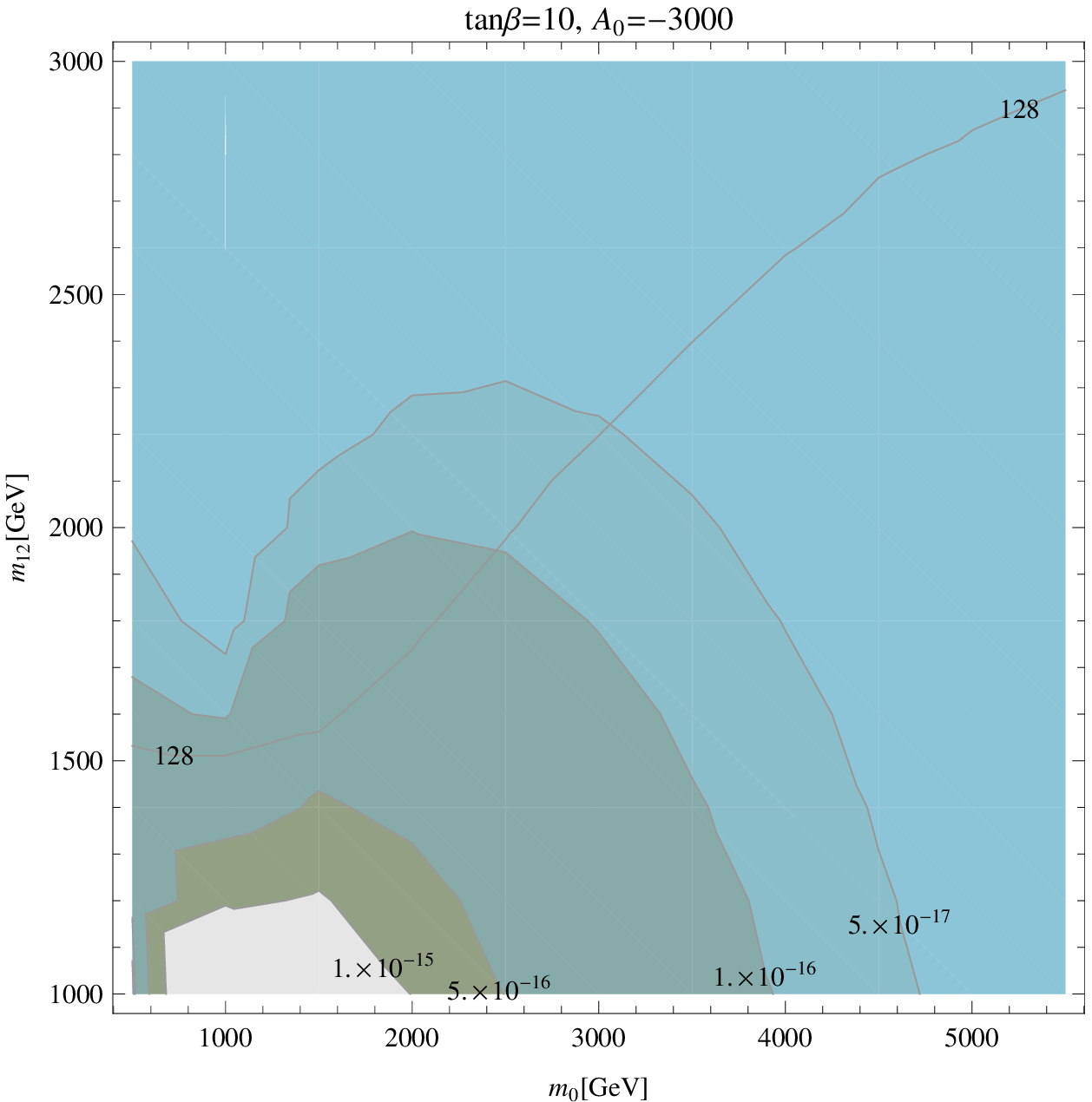  ,scale=0.51,angle=0,clip=}\\
\vspace{0.2cm}
\psfig{file=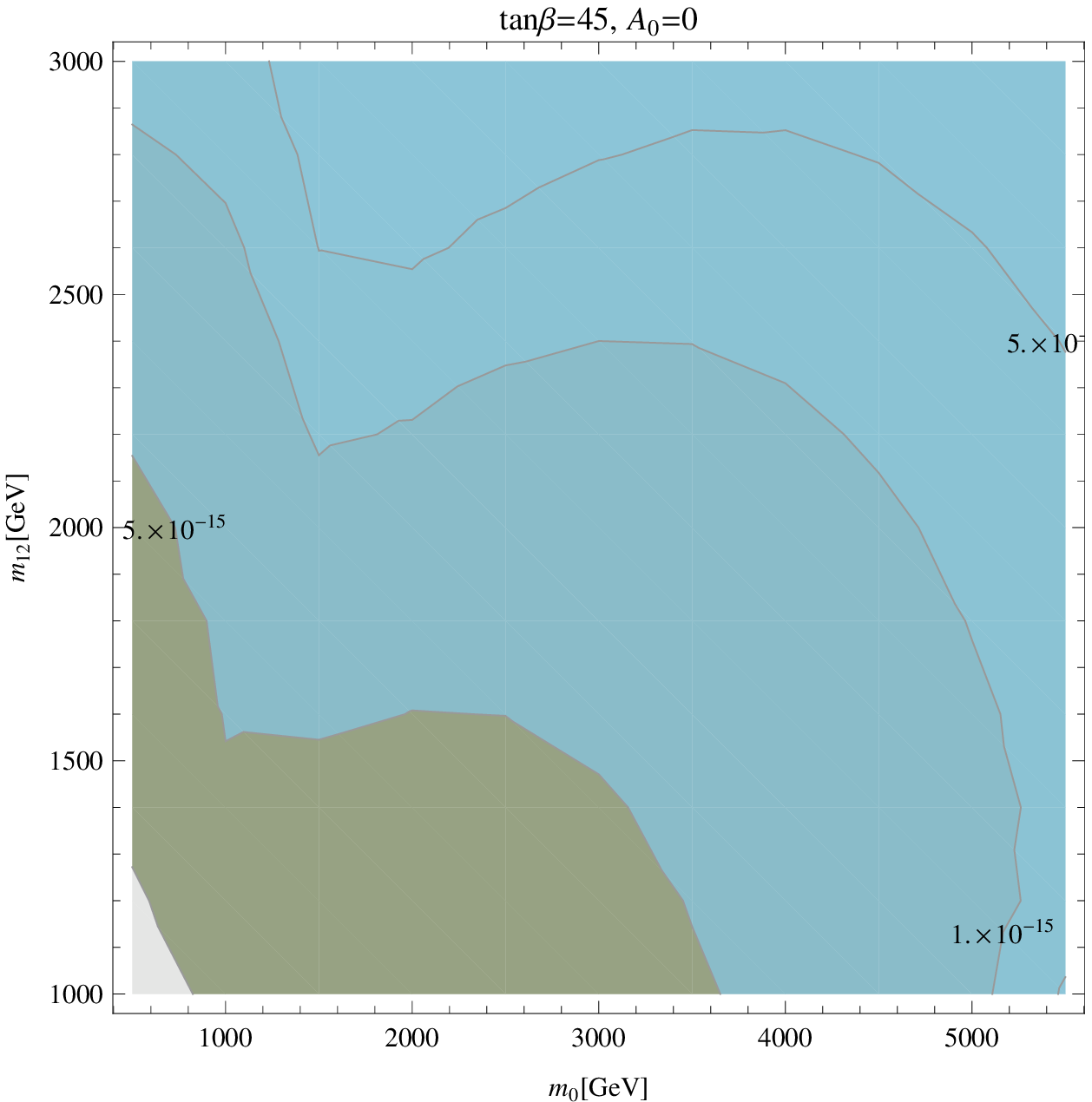 ,scale=0.51,angle=0,clip=}
\psfig{file=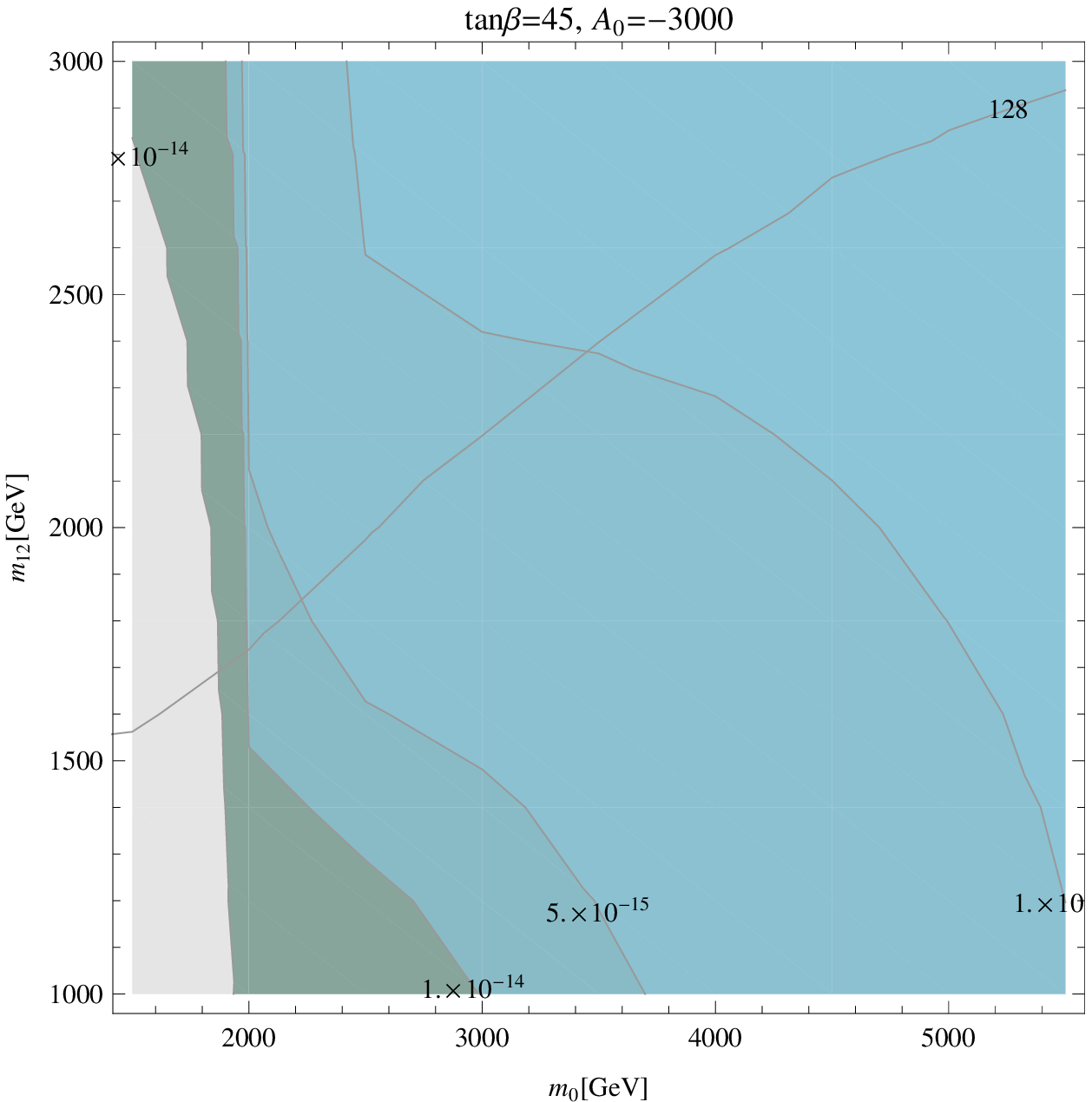   ,scale=0.51,angle=0,clip=}\\
\vspace{-0.2cm}
\end{center}
\caption[Contours of BR($h \to \tau \mu$)  in the
  $m_0$--$m_{1/2}$ plane]{Contours of BR($h \to \tau \mu$)  in the
  $m_0$--$m_{1/2}$ plane for different values of $\tb$ and $A_0$ in
  the \CMSSMI.  Lines labeled as 122 and 128 are as descibred in Figs.\ref{fig:DelLLL23} and \ref{fig:BrmegSSI}}   
\label{fig:HTauMueSSI}
\end{figure}

As we explained before, we do not expect large BR for LFVHD, due to the fact that in our models they are correlated to the restricitive bounds on the cLFV decays. \reffi{fig:HMueESSI} shows the results for BR($h \to e \mu$). The largest value is of the \order{10^{-16}} for low $m_0$ and $m_{1/2}$ values but is excluded from BR($\mu \to e \gamma$). In the allowed range they are typically \order{10^{-18}}. Similarly \reffi{fig:HTauESSI} and \reffi{fig:HTauMueSSI} shows the predictions for BR($h \to e \tau$) and BR($h \to \tau \mu$) respectively. Predictions of the \order{10^{-14}} and \order{10^{-12}} are possible for BR($h \to e \tau$) and BR($h \to \tau \mu$) in the lower left region of the $m_0$--$m_{1/2}$ plane respectively but are excluded from BR($\mu \to e \gamma$) bound. In the allowed region they are of the \order{10^{-16}} or less. These results could not explain a CMS-type excess. We have also analysed other high scale see-saw models like Type II and Type III see-saw. However,  the predictions for LFVHD are again very small compared to a possible  CMS-type excess and we do not show them here. While it is possible to get large predictions for the LFVHD by using neutrino textures that somehow suppress BR($\mu \to e \gamma$), however for the realistic scenerios it is very difficult to get large predictions because off-diagonal enteries in the mass matrix of the slepton are the only source of LFV and large off-diagonal enteries will result in larger values of BR($\mu \to e \gamma$) unless mixing between first and second generation of the leptons is artificially suppressed. Although, ATLAS reports  are not in contradiction with CMS ones,  it remains to be seen how these results will develop with the LHC Run II.


%% file: sec6_conclu.tex
\section{Conclusions}
In this paper we have analyzed the Lepton Flavor Violation effects arising from the supersymmetric extension of the SM. We study several observables that can be sensitive to these effects.  We take into account the restrictions imposed by the non-observation of 
charged Lepton Flavor Violation (cLFV) processes on the MSSM slepton mass parameters to study the impact of LFV effects to lepton flavour violating decays of $\cp$-even light Higgs boson (LFVHD). As a computation framework, we consider first a model independent selection of parameters of the MSSM and later some specific  neutrino motivated SUSY models: constrained MSSM (CMSSM) extended by high scale seesaw models in particular Type-I seesaw mechanism.    

For the model independent approach of \refse{sec:NR_modelInd} we consider six phenomenologically motivated benchmark 
points. These scenarios were studied before to extract the various  $\deFABij$ allowed by cLFV processes in  \citere{Arana-Catania:2013nha}. Here, we impose their values to evaluate decay rates for the LFVHD. It turns out that it is very difficult to get large predictions for the LFVHD due to very strict constraints from cLFV decays. The prediction for $\br(h \to \mu \tau)$ can be \order{10^{-9}} at maximum, which is very small compared to the possible CMS-type excess.    

Going to the \CMSSMI\, the numerical results were presented in \refse{NR-CMSSMI}. We have chosen a set of parameters consistent with   the observed neutrino data and simultaneously 
induces large LFV effects and induces {\em relatively} large corrections
to the calculated observables. Consequently, parts of the parameter
space are excluded by the experimental bounds on $\br(\mu \to e \ga)$.  As it was expected the prediction for the BR of LFVHD turned out very small in all the scenarios considered. We can conclude that we may need additional sources of lepton flavor violation (other then already present in the high scale see-saw models) to explain a CMS-type excess for the channel $\br(h \to \mu \tau)$.  Other neutrino motivated SUSY scenarios such as the inverse see-saw models can enhance lepton flavor violating Higgs boson decay rates \cite{Arganda:2015naa,Hammad:2016bng}. However, the latest results from CMS, if confirmed, will impose severe constraints on these models. 

%% file: LFVHD_mainfile.bbl
\clearpage

\begin{thebibliography}{99} 
%
%

\bibitem{Neutrino-Osc} S.~Fukuda {\em et al.}  [Super-Kamiokande Collaboration],
                       Phys.\ Rev.\ Lett. {\bf 86},  5656 (2001).
                       {\tt arXiv:hep-ex/0103033};
S.~Fukuda {\em et al.}  [Super-Kamiokande Collaboration],
Phys.\ Rev.\ Lett. {\bf 86},  5651 (2001).
{\tt arXiv:hep-ex/0103032};
S.~Fukuda {\em et al.}  [Super-Kamiokande Collaboration],
Phys.\ Lett. B  {\bf 539},  179 (2002).
{\tt arXiv:hep-ex/0205075};
M.~Apollonio {\em et al.}  [CHOOZ Collaboration],
Phys.\ Lett.  B  {\bf 466},  415 (1999).
{\tt arXiv:hep-ex/9907037};
Q.~Ahmad {\em et al.}  [SNO Collaboration],
Phys.\ Rev.\ Lett. {\bf 87},  071301 (2001).
{\tt arXiv:nucl-ex/0106015};
Q.~Ahmad {\em et al.}  [SNO Collaboration],
Phys.\ Rev.\ Lett. {\bf 89},  011301 (2002).
{\tt arXiv:nucl-ex/0204008};
M.~Ambrosio et al. [MACRO Collaboration], 
Phys.\ Lett.  B {\bf 517},  59 (2001);
G.~Giacomelli and M.~Giorgini [MACRO Collaboration], 
{\tt arXiv:hep-ex/0110021};
K. Eguchi {\em et al.} [KamLAND Collaboration],
{\tt arXiv:hep-ex/0212021}

\bibitem{mssm} H.~Nilles, 
               Phys.\ Rept. {\bf 110},  1 (1984);  
               H.~Haber and G.~Kane, 
               Phys.\ Rept.  {\bf 117}, 75 (1985);  
               R.~Barbieri, 
               Riv.\ Nuovo Cim. {\bf 11},  1 (1988) 

\bibitem{Hall:1985dx}
  L.~J.~Hall, V.~A.~Kostelecky and S.~Raby,
  Nucl.\ Phys.\  B {\bf 267}, 415 (1986).

\bibitem{Borzumati:1986qx}
  F.~Borzumati and A.~Masiero,
  Phys.\ Rev.\ Lett.\  {\bf 57} (1986) 961;

\bibitem{cms}
  V.~Khachatryan {\it et al.} [CMS Collaboration],
  Phys.\ Lett.\ B {\bf 749} (2015) 337
  {\tt arXiv:1502.07400 [hep-ex]}.

\bibitem{atlas}
  G.~Aad {\it et al.} [ATLAS Collaboration],
  {\tt arXiv:1604.07730 [hep-ex].}

\bibitem{seesaw:I} P.~Minkowski,
                   Phys.\ Lett.  B {\bf 67}, 421 (1977);
  M.~Gell-Mann, P.~Ramond and R.~Slansky, in {\it Complex Spinors and
Unified Theories} eds.\ P.~Van.~Nieuwenhuizen and D.~Freedman,
{\it Supergravity} (North-Holland, Amsterdam, 1979),
p.315 [Print-80-0576 (CERN)];
%
T.~Yanagida, in {\it Proceedings of the Workshop on the Unified Theory
and the Baryon Number in the Universe}, eds.\ O.~Sawada and
A.~Sugamoto (KEK, Tsukuba, 1979), p.95;
%
S.~Glashow, in {\it Quarks and Leptons}, eds.\ M.~L\'evy {\em et al}. 
(Plenum Press, New York, 1980), p.687;
%
R.~Mohapatra and G.~Senjanovi\'c,
Phys.\ Rev.\ Lett. {\bf 44},  912 (1980)

\bibitem{King:2003jb}
  S.~F.~King,
  Rept.\ Prog.\ Phys.\  {\bf 67} (2004) 107
  {\tt hep-ph/0310204.}

\bibitem{Senjanovic:2011zz}
  G.~Senjanovic,
  Riv.\ Nuovo Cim.\  {\bf 34} (2011) 1.
  doi:10.1393/ncr/i2011-10061-8

\bibitem{LFVhisano}
J.~Hisano, T.~Moroi, K.~Tobe and M.~Yamaguchi,
Phys.\ Rev.  D {\bf 53} 2442 (1996)

\bibitem{gllv}
M.~G\'omez, G.~Leontaris, S.~Lola and J.~Vergados, 
Phys. Rev. D  {\bf 59},  116009 (1999).
{\tt arXiv:hep-ph/9810291}

\bibitem{casas-ibarra}
J.~Casas and A.~Ibarra,
Nucl.\ Phys.  B {\bf 618}, 171 (2001).
{\tt arXiv:hep-ph/0103065}

\bibitem{Mismatch}
J.~Ellis, M.~E.~G\'omez, G.~Leontaris, S.~Lola and D.~Nanopoulos,
Eur.\ Phys.\ J. C  {\bf 14}, 319 (2000). 
{\tt arXiv:hep-ph/9911459}

\bibitem{Antusch}
S.~Antusch, E.~Arganda, M.~Herrero and A.~Teixeira,
JHEP {\bf 0611}, 090 (2006). 
{\tt arXiv:hep-ph/0607263}

\bibitem{Antusch2}
S.~Antusch, J.~Kersten, M.~Lindner, M.~Ratz and M.~A.~Schmidt,
JHEP {\bf 0503} (2005) 024.

\bibitem{Schwieger:1998dd}
  J.~Schwieger, T.~S.~Kosmas and A.~Faessler,
  Phys.\ Lett.\ B {\bf 443} (1998) 7.

\bibitem{Divari:2002sq}
  P.~C.~Divari, J.~D.~Vergados, T.~S.~Kosmas and L.~D.~Skouras,
  Nucl.\ Phys.\ A {\bf 703} (2002) 409
  {\tt [nucl-th/0203066].}

\bibitem{Biggio:2010me}
  C.~Biggio and L.~Calibbi,
  JHEP {\bf 1010} (2010) 037
  {\tt arXiv:1007.3750 [hep-ph].}

\bibitem{Figueiredo:2013tea}
  A.~J.~R.~Figueiredo and A.~M.~Teixeira,
  JHEP {\bf 1401} (2014) 015
  {\tt arXiv:1309.7951 [hep-ph].}

\bibitem{Chowdhury:2013jta}
  D.~Chowdhury and K.~M.~Patel,
  Phys.\ Rev.\ D {\bf 87} (2013) no.9,  095018
  {\tt arXiv:1304.7888 [hep-ph].}

\bibitem{Krauss:2013gya}
  M.~E.~Krauss, W.~Porod, F.~Staub, A.~Abada, A.~Vicente and C.~Weiland,
  Phys.\ Rev.\ D {\bf 90} (2014) no.1,  013008
  {\tt arXiv:1312.5318 [hep-ph].}

\bibitem{Goto:2014vga}
  T.~Goto, Y.~Okada, T.~Shindou, M.~Tanaka and R.~Watanabe,
  Phys.\ Rev.\ D {\bf 91} (2015) no.3,  033007
  {\tt arXiv:1412.2530 [hep-ph]}.

\bibitem{Kersten:2014xaa}
  J.~Kersten, J.~h.~Park, D.~Stöckinger and L.~Velasco-Sevilla,
  JHEP {\bf 1408} (2014) 118
  {\tt arXiv:1405.2972 [hep-ph]}.

\bibitem{Vicente:2015cka}
  A.~Vicente,
  Adv.\ High Energy Phys.\  {\bf 2015} (2015) 686572
  {\tt arXiv:1503.08622 [hep-ph].}

\bibitem{Barbieri:1995tw}
  R.~Barbieri, L.~J.~Hall and A.~Strumia,
  Nucl.\ Phys.\ B {\bf 445} (1995) 219
  {\tt hep-ph/9501334.}

\bibitem{Gomez:1995cv}
  M.~E.~Gomez and H.~Goldberg,
  Phys.\ Rev.\ D {\bf 53} (1996) 5244
  {\tt hep-ph/9510303}.

\bibitem{pedro} 
  M.~E.~Gomez, S.~Lola, P.~Naranjo and J.~Rodriguez-Quintero,
  JHEP {\bf 1006} (2010) 053
  {\tt arXiv:1003.4937 [hep-ph].}
%

\bibitem{Ellis:2016qra}
  J.~Ellis, K.~Olive and L.~Velasco-Sevilla,
{\tt  arXiv:1605.01398 [hep-ph].}
%

\bibitem{Magg:1980ut}
Magg, M. and Wetterich, C.
\newblock {\em Phys. Lett.}{ \bf B94}, 61 (1980).

\bibitem{Lazarides:1980nt}
Lazarides, G., Shafi, Q., and Wetterich, C.
\newblock {\em Nucl. Phys.}{ \bf B181}, 287 (1981).

\bibitem{Foot-Type-III}
Foot, R., Lew, H., He, X.~G., and Joshi, G.~C.
\newblock {\em Z. Phys.}{ \bf C44}, 441 (1989).

\bibitem{Ma:2002pf}
Ma, E. and Roy, D.~P.
\newblock {\em Nucl. Phys.}{ \bf B644}, 290--302 (2002).
%

\bibitem{Brignole:2003iv}
  A.~Brignole and A.~Rossi,
  Phys.\ Lett.\ B {\bf 566} (2003) 217
  {\tt hep-ph/0304081.}

\bibitem{Brignole:2004ah}
  A.~Brignole and A.~Rossi,
  Nucl.\ Phys.\ B {\bf 701} (2004) 3
  {\tt hep-ph/0404211.}

\bibitem{Kanemura:2004cn}
  S.~Kanemura, K.~Matsuda, T.~Ota, T.~Shindou, E.~Takasugi and K.~Tsumura,
  Phys.\ Lett.\ B {\bf 599} (2004) 83
  {\tt hep-ph/0406316.}

\bibitem{Paradisi:2005tk}
  P.~Paradisi,
  JHEP {\bf 0602} (2006) 050
  {\tt hep-ph/0508054.}

\bibitem{Arana-Catania:2013xma}
  M.~Arana-Catania, E.~Arganda and M.~J.~Herrero,
  JHEP {\bf 1309} (2013) 160
   Erratum: [JHEP {\bf 1510} (2015) 192]
  {\tt arXiv:1304.3371 [hep-ph].}

\bibitem{Arganda:2015uca}
  E.~Arganda, M.~J.~Herrero, R.~Morales and A.~Szynkman,
  JHEP {\bf 1603} (2016) 055
  {\tt arXiv:1510.04685 [hep-ph]}.

\bibitem{Aloni:2015wvn}
  D.~Aloni, Y.~Nir and E.~Stamou,
  JHEP {\bf 1604} (2016) 162
  {\tt arXiv:1511.00979 [hep-ph]}.

\bibitem{feynarts} J.~K\"ublbeck, M.~B\"ohm and A.~Denner, 
                   Comput. Phys. Commun. {\bf 60}, 165 (1990);
                   T.~Hahn, 
                   Comput. Phys. Commun. {\bf 140},  418 (2001).
                   {\tt arXiv:hep-ph/0012260}

\bibitem{famssm}   T.~Hahn and C.~Schappacher, 
                   Comput. Phys. Commun. {\bf 143},  54 (2002).
                   {\tt arXiv:hep-ph/0105349}
                   The program and the user's guide 
                   are available via {\tt www.feynarts.de} .

\bibitem{drhoLFV} M.~G{\'o}mez, T.~Hahn, S.~Heinemeyer, M.~Rehman, 
                  Phys.\ Rev.  D {\bf 90}, 074016 (2014).  
                  {\tt arXiv:1408.0663 [hep-ph]}  

\bibitem{Gomez:2015ila}
  M.~E.~Gomez, S.~Heinemeyer and M.~Rehman,
  Eur.\ Phys.\ J.\ C {\bf 75} (2015) no.9,  434
  {\tt arXiv:1501.02258 [hep-ph]}.

\bibitem{Arganda:2004bz}
  E.~Arganda, A.~M.~Curiel, M.~J.~Herrero and D.~Temes,
  Phys.\ Rev.\ D {\bf 71} (2005) 035011
  {\tt hep-ph/0407302.}

\bibitem{Arana-Catania:2013nha}
  M.~Arana-Catania, S.~Heinemeyer and M.~Herrero,
  Phys.\ Rev.  D {\bf 88}, 015026 (2013). 
  {\tt arXiv:1304.2783 [hep-ph]}

\bibitem{Kuno:1999jp} 
  Y.~Kuno and Y.~Okada, 
  Rev.\ Mod.\ Phys. {\bf 73}, 151 (2001). 
  {\tt arXiv:hep-ph/9909265}

\bibitem{DiracNu}
  S.~Bilenky, S.~Petcov and B.~Pontecorvo,
  Phys.\ Lett.  B {\bf 67}, 309 (1977);
  W.~Marciano and A.~Sanda,
  Phys.\ Lett.  B  {\bf 67}, 303 (1977) 

\bibitem{MajoranaNu}
  T.~Cheng, L.-F.~Li,
  Phys.\ Rev.\ Lett. {\bf 45}, 1908 (1980) 

\bibitem{Adam:2013mnn} 
  J.~Adam {\it et al.}  [MEG Collaboration],
  {\tt arXiv:1303.0754 [hep-ex].}

\bibitem{Aubert:2009ag} 
  B.~Aubert {\it et al.}  [BABAR Collaboration],
  Phys.\ Rev.\ Lett.\  {\bf 104}, 021802 (2010)
   {\tt arXiv:0908.2381 [hep-ex]}. 

\bibitem{Bellgardt:1987du}
  U.~Bellgardt {\it et al.}  [SINDRUM Collaboration],
  Nucl.\ Phys.\  B {\bf 299}, 1 (1988)

\bibitem{Bertl:2006up}
  W.~Bertl {\it et al.}  [SINDRUM II Collaboration],
  Eur.\ Phys.\ J.\  C {\bf 47}, 337 (2006).

\bibitem{Hayasaka:2010np}
  K.~Hayasaka, K.~Inami, Y.~Miyazaki, K.~Arinstein, V.~Aulchenko, T.~Aushev, A.~M.~Bakich, A.~Bay {\it et al.},
  Phys.\ Lett.\  {\bf B687 } (2010)  139-143.
  {\tt arXiv:1001.3221 [hep-ex].}

\bibitem{Khachatryan:2016vau} 
G.~Aad {\it et al.} [ATLAS and CMS Collaborations], 
  JHEP {\bf 1608} (2016) 045 

\bibitem{deFlorian:2016spz}
  D.~de Florian {\it et al.} [LHC Higgs Cross Section Working Group Collaboration],
  ``Handbook of LHC Higgs Cross Sections: 4. Deciphering the Nature of the Higgs Sector,''
  {\tt arXiv:1610.07922.}

\bibitem{CMSAad2014} G. Aad {\em et al.} [ATLAS Collaboration], Phys. Rev. Lett. {\bf 112}, 201802 (2014). {\tt arXiv:1402.3244 [hep-ex]}

\bibitem{CMSChatrchyan2014} S. Chatrchyan {\em et al.} [CMS Collaboration], Eur. Phys. J. C {\bf 74}, 2980 (2014). {\tt arXiv:1404.1344 [hep-ex]}

\bibitem{CMS_not}
http://cms-results.web.cern.ch/cms-results/public-results/preliminary-results/HIG-16-005/index.html

\bibitem{hbs2016} M.~G{\'o}mez, S.~Heinemeyer, M.~Rehman, 
                  Phys.\ Rev.  D {\bf 93}, 095021 (2016).  
                  {\tt arXiv:1511.04342 [hep-ph]}

\bibitem{feynhiggs} S.~Heinemeyer, W.~Hollik and G.~Weiglein,
                   Comput. Phys. Commun. {\bf 124}, 76 (2000). 
                   {\tt arXiv:hep-ph/9812320};
                   T.~Hahn, S.~Heinemeyer, W.~Hollik, H.~Rzehak and
                   G.~Weiglein, 
                   Comput.\ Phys.\ Commun. {\bf 180}, 1426 (2009)
                   see {\tt www.feynhiggs.de}

\bibitem{mhiggslong} S.~Heinemeyer, W.~Hollik and G.~Weiglein,
                    Eur. Phys. J.  C {\bf 9}, 343 (1999). 
                    {\tt arXiv:hep-ph/9812472}

\bibitem{mhiggsAEC} G.~Degrassi, S.~Heinemeyer, W.~Hollik,
                   P.~Slavich and G.~Weiglein,
                   Eur. Phys. J.  C {\bf 28},  133 (2003).
                   {\tt arXiv:hep-ph/0212020}

\bibitem{mhcMSSMlong}
                   M.~Frank, T.~Hahn, S.~Heinemeyer, W.~Hollik, 
                   R.~Rzehak and G.~Weiglein,
                   JHEP {\bf 0702},  047 (2007).
                   {\tt arXiv:hep-ph/0611326}

\bibitem{Mh-logresum}  T.~Hahn, S.~Heinemeyer, W.~Hollik, H.~Rzehak and
  G.~Weiglein,
  Phys. Rev. Lett. {\bf 112}, 141801 (2014). 
  {\tt arXiv:1312.4937 [hep-ph]}

\bibitem{Haber:1989xc} 
  H.~Haber and Y.~Nir,
  Nucl.\ Phys.  B {\bf 335}, 363 (1990) 

\bibitem{LHCHiggs}
  S.~Chatrchyan {\it et al.}  [CMS Collaboration],
  arXiv:1303.4571 [hep-ex];\\
 Pedrame Bargassa, talk given at ``Rencontres de Moriond EW 2014'',\\
{\tt
https://indico.in2p3.fr/getFile.py/access?contribId=189\&sessionId=0\\
 \&resId=1\&materialId=slides\&confId=9116};\\
 Mike Flowerdew, talk given at ``Rencontres de Moriond EW 2014'',\\
{\tt
https://indico.in2p3.fr/getFile.py/access?contribId=169\&sessionId=0\\
 \&resId=0\&materialId=slides\&confId=9116};\\
 Paul Thompson, talk given at ``Rencontres de Moriond EW 2014'',\\
{\tt
https://indico.in2p3.fr/getFile.py/access?contribId=220\&sessionId=8\\
\&resId=0\&materialId=slides\&confId=9116};\\
Kevin Einsweiler, talk given at ``Rencontres de Moriond EW 2014'',\\
{\tt
https://indico.in2p3.fr/getFile.py/access?contribId=227\&sessionId=1\\
\&resId=1\&materialId=slides\&confId=9116}~.

\bibitem{higgsbounds} P.~Bechtle, O.~Brein, S.~Heinemeyer, G.~Weiglein
  and K.~Williams, 
  Comput.\ Phys.\ Commun. {\bf 181}, 138 (2010). 
  {\tt arXiv:0811.4169 [hep-ph]};
  Comput.\ Phys.\ Commun. {\bf 182}, 2605 (2011). 
  {\tt arXiv:1102.1898 [hep-ph]};
P.~Bechtle, O.~Brein, S.~Heinemeyer, O.~St{\aa}l, T.~Stefaniak, G.~Weiglein
and K.~Williams,
  Eur.\ Phys.\ J. C {\bf 74}, 2693 (2014). 
  {\tt arXiv:1311.0055 [hep-ph]}

\bibitem{Babu:2002et}
  K.~S.~Babu and C.~Kolda,
  Phys.\ Rev.\ Lett.\  {\bf 89} (2002) 241802
  {\tt hep-ph/0206310.}

\bibitem{Dedes:2002rh}
  A.~Dedes, J.~R.~Ellis and M.~Raidal,
  Phys.\ Lett.\ B {\bf 549} (2002) 159
  {\tt hep-ph/0209207.}

\bibitem{Cannoni:2008bg}
  M.~Cannoni and O.~Panella,
  Phys.\ Rev.\ D {\bf 79} (2009) 056001
  {\tt arXiv:0812.2875 [hep-ph].}

\bibitem{Hisano:2010es}
  J.~Hisano, S.~Sugiyama, M.~Yamanaka and M.~J.~S.~Yang,
  Phys.\ Lett.\ B {\bf 694} (2011) 380
  {\tt arXiv:1005.3648 [hep-ph]}.
\cite{CMS_not}

\bibitem{a-t-t}
The ATLAS Collaboration, ATLAS-CONF-2016-085. \\
The CMS Collaboration, CMS-PAS-HIG-16-037.

\bibitem{Cannoni:2013gq}
  M.~Cannoni, J.~Ellis, M.~G\'omez and S.~Lola,
  Phys.\ Rev. D  {\bf 88},  075005 (2013).
  {\tt arXiv:1301.6002 [hep-ph]}

\bibitem{EGL}
J.~Ellis, M.~G\'omez and S.~Lola,
JHEP {\bf 0707},  052 (2007).
{\tt arXiv:hep-ph/0612292}

\bibitem{Porod:2003um} W.~Porod, 
  Comput. Phys. Commun. {\bf 153},  275 (2003).
  {\tt arXiv:hep-ph/0301101}

\bibitem{SLHA} P.~Skands et al.,
  JHEP {\bf 0407}, 036 (2004). 
  {\tt arXiv:hep-ph/0311123};
  B.~Allanach et al.,
  Comput.\ Phys.\ Commun. {\bf 180}, 8 (2009). 
  {\tt arXiv:0801.0045 [hep-ph]}

\bibitem{Olive:2016efh}
  K.~A.~Olive,
  PoS DSU {\bf 2015} (2016) 035
  {\tt arXiv:1604.07336 [hep-ph]}.

\bibitem{Arganda:2015naa}
  E.~Arganda, M.~J.~Herrero, X.~Marcano and C.~Weiland,
  Phys.\ Rev.\ D {\bf 93} (2016) no.5,  055010
  {\tt arXiv:1508.04623 [hep-ph].}

\bibitem{Hammad:2016bng}
  A.~Hammad, S.~Khalil and C.~S.~Un,
 {\tt arXiv:1605.07567 [hep-ph].}

\end{thebibliography}
